\documentclass[preprint]{aastex}

\newcommand{\HI}{H\,{\small I}}
\newcommand{\HII}{H\,{\sc ii}}
\newcommand{\Mo}{\mbox{M$_{\odot}$}}
\newcommand{\Hz}{\mbox{H$_0= 75$~km~s$^{-1}$~Mpc$^{-1}$}}
\newcommand{\BV}{\mbox{$\mbox{B} - \mbox{V}$}}
\newcommand{\x}{\mbox{$\times$}}
\newcommand{\Ha}{\mbox{H$_{\alpha}$}}
\newcommand{\Hb}{\mbox{H$_{\beta}$}}
\newcommand{\kms}{\mbox{km~s$^{-1}$}}
\newcommand{\Lfir}{\mbox{$L_{\mbox{\tiny FIR}}$}}
\newcommand{\Lb}{\mbox{$L_{\mbox{\tiny B}}$}}
\newcommand{\MHH}{\mbox{$M_{\mbox{\tiny H$_2$}}$}}
\newcommand{\HH}{\mbox{H$_2$}}
\newcommand{\Mb}{\mbox{$M_{\mbox{\tiny B}}$}}
\newcommand{\cmm}{\mbox{cm$^{-2}$}}
\newcommand{\VK}{\mbox{$\mbox{V} - \mbox{K}$}}
\newcommand{\sbu}{${\rm mag\,\,arcsec^{-2 }}$}
\newcommand{\sbV}{\mbox{$\mu _{\mbox{\tiny V}}$}}
\newcommand{\sbB}{\mbox{$\mu _{\mbox{\tiny B}}$}}
\newcommand{\Ab}{\mbox{$A_{\mbox{\tiny B}}$}}
\newcommand{\LHa}{\mbox{$L_{\mbox{\tiny H}\alpha}$}}
\newcommand{\OIIIc}{\mbox{[OIII]$_{\lambda 4363}$}}

\newcommand{\OII}{\mbox{[OII]$_{\lambda 3727}$}}

\newcommand{\OIIIt}{\mbox{[OIII]}}
\newcommand{\OIIt}{\mbox{[OII]}}
\newcommand{\NIIt}{\mbox{[NII]}}

\newcommand{\NIIb}{\mbox{[NII]$_{\lambda 6584}$}}
\newcommand{\SIIt}{\mbox{[SII]}}

\newcommand{\usbflux}{\mbox{erg~s$^{-1}$~cm$^{-2}$~arcsec$^{-2}$}}
\newcommand{\Zo}{\mbox{Z$_{\odot}$}}
\newcommand{\NHI}{\mbox{$N_{\mbox{\tiny HI}}$}}
\newcommand{\tabsp}{\noalign{\smallskip}}
\newcommand{\Lo}{\mbox{L$_{\odot}$}}
\newcommand{\er}{\mbox{$\pm$}}

\newcommand{\ufluxm}{\mbox{erg~cm$^{-2}$~s$^{-1}$~\AA$^{-1}$}}
\newcommand{\mJyb}{\mbox{mJy~beam$^{-1}$}}

\slugcomment{Revised version,  19/05/2000}
\bibpunct{(}{)}{,}{a}{}{,}

\shorttitle{Formation of a tidal dwarf galaxy in Arp~245}
\shortauthors{Duc et al.}

\begin{document}

\title{Formation of a tidal dwarf galaxy in the interacting system Arp~245 (NGC~2992/93)}

\author{P.--A. Duc\altaffilmark{1,2}, E. Brinks\altaffilmark{3}, V. Springel\altaffilmark{4,5},
  B. Pichardo\altaffilmark{6}, P. Weilbacher\altaffilmark{7} and I.F. Mirabel\altaffilmark{8,9}}

\altaffiltext{1} {CNRS URA 2052 and CEA--Saclay, DSM, DAPNIA, Service d'Astrophysique, 91191 Gif--sur--Yvette
 Cedex, France; email: paduc@cea.fr}
\altaffiltext{2}  {University of Cambridge, Institute of Astronomy, Madingley Road, 
Cambridge, CB3~0HA, UK}
\altaffiltext{3} {Departamento de Astronom\'{\i}a, Apdo. Postal 144, Guanajuato, Gto. 36000, Mexico}
\altaffiltext{4} {Max--Planck--Institut f\"ur Astrophysik, Karl--Schwarzschild--Strasse 1, 85740 Garching bei M\"unchen, Germany}
\altaffiltext{5} {Harvard Smithsonian Center for Astrophysics, 60
Garden Street, MS51, Cambridge, MA 02138, USA}
\altaffiltext{6} {Instituto de Astronom\'{\i}a, UNAM, Mexico}
\altaffiltext{7} {Universit\"atssternwarte, Geismarlandstr.~11, D--37083 G\"ottingen, Germany}
\altaffiltext{8} {CEA--Saclay, DSM, DAPNIA, Service d'Astrophysique, 91191 Gif--sur--Yvette Cedex, France}
\altaffiltext{9} {IAFE, cc 67, snc 28 (1948), Buenos Aires, Argentina}

\begin{abstract}
Among the various phenomena observed in interacting galaxies is  the
 ejection due to tidal forces
 of stellar and gaseous material into the intergalactic medium and  its subsequent 
rearranging which can lead to the formation of self--gravitating tidal dwarf galaxies 
 (TDGs). We investigate this process with  a detailed 
 multiwavelength study of the interacting system Arp~245 and a numerical model 
of the collision computed with a Tree--SPH code. Our observations consist of
 optical/near--infrared broad
band imaging, H$\alpha$ imaging, optical  spectroscopy,  
\HI\ VLA cartography and CO line mapping. The system, composed of the two spiral galaxies
NGC~2992 and NGC~2993, is
 observed at an early stage of the interaction, about 100 Myr
after perigalacticon, though at a time when tidal tails have already developed. 
 The VLA observations disclose a third partner to the interaction: an edge--on,
flat galaxy, FGC~0938, which looks strikingly undisturbed and might just be falling
towards the NGC 2992/93 system. 
 Our \HI\ map shows prominent counterparts to the optical  tails.
Whereas the stellar and gaseous components of the plume that
 originates from NGC~2992 match, the stellar and \HI\ tails
emanating from NGC~2993  have a different morphology. In particular,
 the \HI\ forms a ring, a feature that has been successfully
 reproduced by our numerical
simulations. The \HI\ emission in the system as a whole peaks at the tip of
the NGC~2992 tail where a gas reservoir of about $10^9~\Mo$, about
60\% of the \HI\ towards NGC~2992, coincides with
a star--forming optical condensation,  A245N.
  The latter tidal object  exhibits properties 
 ranging between those of dwarf irregular galaxies  (structural parameters,
  gas content, star formation rate) and those of spiral disks (metallicity, star
 formation efficiency, stellar population).  Although it is likely,
based on our analysis 
of the HI and model datacube, that A245N might become an independent
dwarf galaxy, the dynamical evidence is still open to debate. Prompted
by the questions raised for this particular object,
 we discuss some issues related to the definition and identification of
TDGs and highlight some 
specific conditions which seem required  to form them. We finally  outline
 what is needed in terms of future numerical simulations in order to further
 our understanding of these objects.

\end{abstract}


%

\section{Introduction}

\nocite{Malphrus97}
\nocite{vanderHulst79}

Research activity in the field of interacting galaxies has increased
 quite dramatically 
 over the last thirty years \cite[see the recent very comprehensive review by][]{Struck99}.
 Galactic collisions  trigger a number of phenomena,
such as inward--transportation  of gas from distances of up to
 kiloparsecs to the nucleus which is thought to be an efficient means to fuel
a central starburst or  nuclear activity.
  The inverse process is the ejection of
material into the intergalactic medium by tidal forces.
 The prominent tidal tails and bridges that emanate from interacting
 galaxies have proved to be important tools to study
 the interaction, constraining the orbital parameters \citep{Toomre72}
 of the collision.  Recently, attempts have even been made to use the
 formation of tidal tails as a diagnostic of the mass distribution of halos within the
 framework of Cold Dark Matter cosmologies
 \citep{Dubinski96,Springel99}. Much less attention has been paid to
 what goes on within these tidal features \citep[see for
 instance][]{Schombert90,Wallin90,Hibbard00}.

Detailed \HI\ maps of a 
number of  interacting systems \citep[e.g.,][ and references therein]{Hibbard96,Kaufman97} have  shown 
 that  a large fraction of the gaseous component of colliding galaxies can be  expelled
 into the galactic halos or even into the intergalactic 
medium as a result of tidal forces. In some  systems, up to 90\% of the atomic
 hydrogen is observed outside the optical disk \citep[like in Arp~105, ][]{Duc97b}. Even if
part of this gas  falls back towards its progenitors  \citep{Hibbard95b},
  a significant amount of gas will be lost for the merger remnant for  time scales of at least 
1--10 Gyr.  The stellar/gaseous tidal debris might be dispersed in the intergalactic/intracluster 
medium where the stellar component then contributes to the diffuse background  light 
 observed in clusters \citep{Gregg98} or recondense within the halo of
the merger and form a new 
generation of galaxies: 
 the so--called  tidal dwarf galaxies \citep[TDGs; see the review by ][]{Duc99b}. 
TDGs are typically found at the tip of tidal tails at distances between 20 and 100 kpc from the 
merging objects, of which at least one should be a gas--rich galaxy. They are gas--rich objects
 that can be as massive as the 
Magellanic Clouds, form stars at a rate which might be as high as in blue compact dwarf galaxies 
(BCDGs) and seem dynamically independent from their parent galaxies.
Although the observational
 evidence for the existence of recycled galaxies has been well
established by now, their formation process
is as yet not well understood. In particular it is not known when and
under which conditions TDGs form during the 
interaction. This is one of the topics of the current paper.

Dating events is a general problem in  extragalactic astronomy that may however be more easily
achieved for interacting systems via numerical simulations.
 The comparison of morphological and/or
kinematical features with predictions based on numerical simulations provides strong indications
 on the age of the collision and therefore sets a constraint on the formation history of TDGs.
Interacting galaxies observed just after the first perigalacticon are particularly attractive. 
Whereas they have already developed  long tails and bridges, 
 their disks are still clearly separated and hence their intrinsic properties (i.e., orientation, 
 sense and amplitude of their rotation, which are key input parameters to the numerical
models) can be well defined. In contrast, galaxies on course for their
first interaction do not show strong perturbations before reaching perigalacticon and
hence do not provide  a simple handle on any time scale. On the other hand, evolved
mergers have lost all memory of the initial
properties.

Arp~245, an interacting system consisting of two spiral galaxies,
 NGC~2992 and NGC~2993,
 appears to be an interesting testcase as it can be fairly easily modelled.  Moreover, it
is a relatively nearby system, at an adopted distance of 31~Mpc\footnote{We use in this
paper \Hz = 100~h~km~s$^{-1}$~Mpc$^{-1}$. At the distance of Arp~245, 1$\arcmin$ corresponds to 9
 kpc}. Its prominent tidal tails host a tidal dwarf galaxy  candidate which,
 because of its proximity, can be studied in detail.
 Finally Arp~245 has already been well studied. NGC~2992, in particular,
 has been the object of numerous articles focusing mostly on its active, Seyfert~1.9 nucleus.
  Radio continuum maps
 have revealed a striking pair of loops near the nucleus with a ``Figure--8'' shape 
\citep{Ulvestad84} which has later on been detected at other
 wavelengths as well \citep{Wehrle88,Chapman00}.
 \citet{Glass97} has monitored the AGN in the near--infrared and has reported an outburst. 
 \citet{Durret87,Durret88,Colbert96a} and \citet{Allen99} have studied
 the ionization cone extending from the AGN and
the surrounding extended emission line
regions. Evidence for outflow has been claimed by \citet{Colbert96a} and \citet{Marquez98}.
NGC~2992 has been observed at many wavelengths from the X--ray regime, where it is a
strong emitter \citep{Marshall81,Gilli00}, to the centimetre wavelengths \citep{Condon82}.  Single
 dish \HI\ \citep{Mirabel84} and CO \citep{Sanders85} data are also available. NGC~2993 
is mentioned in several catalogs of starburst galaxies but has not yet been the subject
of any individual study. Surprisingly, the fact that both galaxies are
 partaking in a spectacular interaction has largely been ignored, apart from
the photometric work by \citet{Schombert90}.

We decided to subject Arp~245 to a comprehensive multiwavelength study. In this first paper, we 
 detail the observations, present
 in particular the first complete \HI\ map of the system, and assess the status of the interaction with
 the help of numerical  N--body/hydrodynamical  simulations. We then focus on
  the tidal features,  emphasizing the properties of the tidal dwarf galaxy  candidate.
Based on these results, and of similar such systems from the
 literature, we attempt to better understand the formation process of TDGs. 
In a second paper (Paper~II, in prep.), we will concentrate on the internal properties of NGC~2992,
 and study in particular its Active Galactic Nucleus (AGN) and the
 ionization filaments.

\section{Observations}
Table~\ref{tab:log} and \ref{tab:VLA} summarize our multi--wavelength observations\footnote{Partly carried out at
the European Southern Observatory, La Silla, Chile (ESO No 54.A--0606, 56.A--0757, 64.N--0163 and 64.N--0361)} 
of Arp~245 and lists technical details. In the next sections we will
discuss these observations in turn.

\subsection{Optical and near--infrared broad--band imaging}

Optical broad band  BVR images of Arp~245
  have been collected in February 1995 with the 3.5--m NTT at la Silla observatory. The red
 arm of EMMI has been used. The weather conditions were photometric and the seeing varied between 
$0\farcs 9$ and $1\farcs 2$.
The $9\farcm2 \x 8\farcm7$ field of view of the R--band image covered the entire interacting system
 including the tidal features whereas the B\footnote{The B--like filter, Bb, optimized for the red arm of
EMMI, has actually been used.} and V band images were somewhat offset
  and missed a small part of the southern tail. However,
B and V images of this part of the system had been previously taken  with the PUMA camera on the
CFHT\footnote{The Canada--France--Hawaii Telescope is operated by the National Research Council of
 Canada, the Centre National de la Recherche Scientifique de France and the University of Hawaii. }
 in February 1992. The weather conditions at CFHT were poor and the seeing was 1$\arcsec$.
Finally  BVR images of a field situated to the South-West of NGC~2993  were obtained in
January 2000 with the ESO 3.6m telescope. The weather conditions were photometric and the seeing 
was 1$\arcsec$. Images from the three telescopes were  eventually combined.
 Landolt fields of photometric standard stars \citep{Landolt92} 
were observed for flux calibration.

Near--infrared JHK$^\prime$ images 
 of NGC~2992 were obtained in February 1996 with the IRAC2B camera installed on the 
MPG/ESO 2.2m. The field of view of each individual image was $2\farcm1 \x 2\farcm1$. Sky images 
were taken of adjacent fields offset by 2$\arcmin$. The entire field  of NGC~2992, including 
the tidal tail, was covered with a mosaic technique. The highest  effective integration time 
 was reached towards the  tail which was observed for a total of 15 min in J, 
12 min in H and 13 min in K$^\prime$. Several photometric standard
stars from the list of UKIRT faint IR standards \citep{Casali92} were observed under photometric
conditions. The  seeing varied between $1\farcs 3$ and $1\farcs 5$.

The reduction of the optical/NIR data has been performed within IRAF with 
standard tasks from the {\sc ccdred} package complemented by  self--written scripts 
 which perform a semi--automatic  processing of the NIR data set. 
An astrometric correction achieving a precision of $\sim
0\farcs 3$ was performed to each optical/NIR image using the positions of several tens of  stars  
 from the USNO A1.0 astrometric catalog \citep{Monet96} queried 
via the ESO {\sc skycat} browser. The frames were corrected for distortions during this process. 
They were registered, PSF--matched and combined to produce color maps.
 An optical ``true color'' image of the system was constructed from the combination of the
 NTT/CFHT/3.6m B, V, and R images. It is shown in Figure~\ref{fig:BVR}. The monochromatic V--band image
 is displayed in Figure~\ref{fig:VHI}. 

Photometry was carried out on the registered images with the {\sc digiphot}
 package in IRAF. Polygonal apertures were chosen to enclose respectively the disks of NGC~2992 and
NGC~2993, their outer regions only, the bridge between both systems, the tidal tails, and the tidal
 dwarf galaxy.
 Because of the complex geometry  of the system, the sky level
 was measured manually and averaged at numerous positions surrounding the  
objects. This procedure was chosen so as to ensure that sky contamination
 by stars or galaxies would be minimal. 
The photometric accuracies take into account variations in the sky background as well as
photon noise. The magnitudes, corrected for 
Galactic extinction are listed in Tables~\ref{tab:partners}, \ref{tab:tidal}, and  \ref{tab:TDG}.
Circular aperture photometry of both NGC~2992 and NGC~2993 was
 performed as well, for comparison
 purposes.  In the optical, our V and R magnitudes  differ by less than 
0.05~mag with data in the literature \citep{Prugniel98} whereas in the B--band, the agreement
 is not as good. The B magnitudes were systematically too faint by 0.1~mag. A scatter in the color terms in
 the blue is expected since the EMMI Bb filter which we have used slightly differs from a
Bessel B filter. As the system does not show large variations in color,
we have applied to all  B--band magnitudes  a correction corresponding
 to an offset of 0.1~mag. This post--calibration has been later on 
validated using the B--band photometric measurements obtained in January 2000. 
 In the near--infrared, comparisons are more  difficult due to the intrinsic
variability of the nuclear flux.  \citet{Glass97} monitored the
AGN and measured excursions of the central K--band flux of up to  0.6 mag due to outbursts.
 Colors appear to be less affected though and our J--H and H--K$^\prime$ colors are within
 0.1~mag of those of  \citet{Alonso-Herrero98}. On average our optical/NIR photometry  might be affected by 
systematic errors of up to 0.1~mag. 
 Note that the instrumental errors quoted in Tables~\ref{tab:tidal} and  
\ref{tab:TDG} do not take this into account.

\subsection{\Ha\ imaging}
An image with a narrow--band filter centered on the redshifted  \Ha\ line has been 
obtained with EMMI during the NTT run. The filter which has a width (FWHM) of 66\AA\ 
 includes light from \Ha\ and \NIIt\ . 
The image covers all of NGC~2992, including its tidal plume and 
 bridge, and the disk of NGC~2993 with part of its tidal tail. Only the southern most region 
has been missed. The exposure time was 15 min.
The continuum was obtained by using a similar
narrow--band filter but offset in wavelength. A halftone \Ha\ image is displayed
 in Figure~\ref{fig:HaHI} and intensity contours of the line emission towards NGC~2993
and towards the northern tidal tail are shown in Figure~\ref{fig:N2993:VHa} and
Figure~\ref{fig:TDG:Ha_spec}, respectively. A contour map towards NGC~2992 will be
presented in Paper~II.
 Our \Ha\  image shows a number of structures 
not seen in the maps previously published \citep[see e.g.][for some
recent ones]{Colbert96a,Allen99}, such as long ionization filaments outside the disk of NGC~2992
 and \HII\ regions at the tip of the northern tidal plume.
Flux calibration was achieved by observing
 a spectroscopic standard star from the list of \citet{Hamuy92} observed with same setup. 
Our integrated flux towards NGC~2992 agrees to within 20\% 
with the flux measured by \citet{Colbert96a}.

\subsection{Optical  spectroscopy}
Longslit spectroscopic observations were carried out in February 1995 with the NTT and EMMI. The grism used
 has an intrinsic resolution of 9\AA. Coupled with the red--arm CCD, it covers the  wavelength range
 3800--8400 \AA. Two spectra were obtained along an axis roughly parallel to the morphological
  axis of NGC~2992. The position
angles were $ -18^\circ$ and $-20.2^\circ$. The $1\farcs 5$ wide
slits encompassed the western regions of 
NGC~2992 and two of the brightest clumps of the tidal dwarf galaxy at the tip of the plume.
The total integration time was 45 min divided in three exposures. The spectra of
several spectroscopic standard stars from the list of \citet{Hamuy92} were measured
 through a wide slit. Weather conditions were photometric.
A spectroscopic follow--up was performed in January 2000 with the EFOSC2 instrument
installed on the ESO 3.6m telescope. The MOS mode offered by punched masks was used
to obtain the optical spectra of several extended objects near NGC~2992 as well as some new
condensations in its tidal dwarf. The grism used had a wider wavelength coverage but a slightly
lower spectral resolution. Five exposures were obtained and the total integration time
was 100 min.

Data reduction, wavelength and flux calibrations were performed with the {\sc twodspec} package
 within IRAF following standard procedures.
Line fluxes and errors were measured with the IRAF {\sc splot} task.  
Data were corrected for extinction using  the
formula:
\[ \frac{I(\lambda)}{I(\Hb)} = \frac{F(\lambda)}{F(\Hb)} * 10^{c*f(\lambda)} \]
where $F(\lambda)$ is the observed line flux, $f(\lambda)$, the reddening function
taken from \citet{Torres-Peimbert89} and
 $c$ the logarithmic reddening correction at \Hb~ obtained from a constant
 $\Ha/\Hb$ Balmer decrement of 2.85.
The lines were not corrected for  underlying stellar absorption.
Table~\ref{tab:TDG:spec} displays the values of
 $c$, the absolute flux and equivalent width of the \Hb~ line and the observed/corrected
 fluxes relative to \Hb~ of the principal lines. 
The accuracy of the absolute flux calibration has been checked using our \Ha\ narrow--band
image. The integrated \Ha\ fluxes of several emission line regions extracted from the \Ha\ image and 
from the slit spectra agree to within 5\% to 15\%.  

\subsection{VLA \HI\ observations}
We obtained observations with the Very Large Array (VLA)\footnote{The
VLA is operated by the National Radio Astronomy Observatory, a
facility of the National Science Foundation operated under cooperative
agreement by Associated Universities, Inc.} of Arp~245 in the 21--cm line of
neutral hydrogen (\HI) in September  1997, using the CS--array\footnote{This
configuration results in a resolution equivalent to that of C--array,
but has some antennae placed at locations usually employed for
D--array, with the aim to provide sufficient short spacing information
so that separate D--array observations are not necessary.}. We used the
calibrator 1331+305 for absolute flux calibration and 0902--142
as secondary calibrator. We determined a flux density for 0902--142 of
$2.89 \pm 0.01$ Jy. The data were taken in correlator mode 4 with two
passbands (or IFs) measuring both right and left hand
polarization. Each band contained 32 channels, and we used an 8--channel
overlap between the bands. The first and last three channels of each
band were discarded because of the decreasing gain toward the edges of
the bandpass. Full details of the observations are presented in Table~\ref{tab:VLA}.

Data reduction and calibration were performed using the Astronomical
Image Processing System ({\sc aips}) package, following standard
procedures. The uv--data from each of the two IFs were calibrated
separately. The visibilities were inspected and bad data points due to
interference were removed. Because of the presence of strong solar
interference at the lowest spatial frequencies, we excluded baselines
shorter than 1 k$\lambda$ (or about 210 m) in the calibration. After
the standard amplitude and phase calibration we went through one cycle
of self--calibration (phase only), improving the final quality of the
images. As reference source we used the bright nuclear emission of
NGC\,2992 which has a flux density of $190\pm2$ mJy.

For each IF we generated ``dirty'' data cubes of line plus continuum
emission. After running several tests we decided that the best maps
were those made with the task {\sc imagr} using the default values for
the {\sc robust} weighting scheme and including the entire observed
uv--range. The first 15 and the last 15 channels of IF1 and
IF2 respectively, were found to be free of line emission. From each
corresponding IF the average of those line--free channels was subtracted to
construct data cubes with line emission only. An average of  all
line free channels from both IFs was used to construct a map of the continuum
emission. This map, after cleaning, is shown in Figure~\ref{fig:rad}. Besides the
nuclear emission due to NGC\,2992 we see emission corresponding to the
nucleus of NGC\,2993. This emission is resolved and measures
$12.2^{\prime\prime} 
\times 9.4^{\prime\prime}$ at a position angle of $58^\circ$. Its peak flux
corresponds to 49 mJy. 

After continuum subtraction the maps were inspected and cleaned, and
the two IFs were appropriately merged.  In order to isolate the \HI\
emission in the maps from the noise, the cube, which has originally a
resolution of $19.4^{\prime\prime} \times 14.4^{\prime\prime}$ was convolved to a circular beam of
$35^{\prime\prime}$. We clipped this cube at a level of twice the rms noise. We
inspected all features in this cube, retaining only those
regions which show emission in at least five consecutive channels.  We
then went back to the original, high resolution cube and applied
conditional blanking, retaining emission in the high resolution cube
from only those regions which were preserved in the processed
$35^{\prime\prime}$
cube. As a fringe benefit this process removed most if not all traces
of solar interference.
Besides emission, \HI\ absorption is seen towards the strong
nuclear source in NGC\,2992.  We ensured that this was preserved in
the conditional blanking process.  In addition to a blanked cube at
the highest resolution ($19.4^{\prime\prime} \times 14.4^{\prime\prime}$) we produced lower
resolution cubes as well, such as at $25^{\prime\prime} \times
25^{\prime\prime}$ and $30^{\prime\prime} \times 30^{\prime\prime}$. As
a last step we obtained the moments of the cubes. The conversion
factors from mJy\,beam$^{-1}$ to Kelvin brightness temperatures are
listed in Table~\ref{tab:VLA}. 

Figure~\ref{fig:HIchannel} displays a mosaic of the \HI\ channel maps at a resolution
of 25$\arcsec$, showing every second channel, superimposed on an image obtained
from the DSS of the same area.  The final \HI\ cube consists of 32
channels with velocities ranging from 2655~\kms\  to
2006~\kms\ with a channel spacing of 21~\kms. Emission is found from 
2592~\kms\ to 2048~\kms. Starting at 2267~\kms\ and extending over some 200~\kms\ to
 2477~\kms\ we see strong absorption against
the nuclear continuum source of NGC\,2992. 

Maps of the integrated \HI\ distribution at a resolution of
 $25^{\prime\prime}$ overlayed on a V--band 
 and on an \Ha\ image of the system are  shown in Figure~\ref{fig:VHI}
 and in Figure~\ref{fig:HaHI}, respectively. In addition to the two main galaxies,
\HI\ emission is clearly associated with the tidal
dwarf galaxy to the north of NGC\,2992, at the tip of the tidal
arm. It is further traced along the northern tidal arm and in the
bridge between NGC\,2992 and NGC\,2993. Surprisingly, there is a huge
ring--like \HI\ structure extending to the southeast of NGC\,2993. 
In addition to emission from the NGC\,2992/93 system, a strong signal
is picked up from what turns out to be a companion object, known as FCG 0938 and located to
the southwest. 
Total \HI\ fluxes of the galaxies are 18.6 Jy\,km\,${\rm s}^{-1}$ for
the interacting system, corresponding to an \HI\ mass of $4.2 \x
10^9~\Mo$. For the companion we find 3.1
Jy\,km\,${\rm s}^{-1}$ which corresponds to an \HI\ mass of $7.0 \x 10^8~\Mo$.

The VLA integrated \HI\ spectrum of NGC~2992 matches well, both qualitatively and
quantitatively, with the single dish \HI\ spectrum obtained by \citet{Mirabel84} at Arecibo.
After beam corrections, the fluxes differ by less than 10~\%.

\subsection{CO observations}
CO(1--0) and CO(2--1) spectra towards the northern tidal tail of Arp~245 were obtained with
 the IRAM 30~m antenna in June 1999. These observations are presented in \citet{Braine00}. 
The CO emission over the entire system, including NGC~2992, NGC~2993, and the tidal features,
 has subsequently been mapped in November 1999  with the SEST 15--m antenna at la Silla. Details regarding 
 those observations will be given in Paper~II.

\section{Results}
\subsection{Status of the interaction}
\subsubsection{The galaxies involved}
Table~\ref{tab:partners} summarizes the global properties of all
members of the system involved in the interaction:
the two spirals of moderate luminosity, NGC~2992 and NGC~2993, and the dwarf galaxy FGC~0938.

The two galaxies NGC~2992
 and NGC~2993  have radial velocities which differ by less than
 100~\kms\ and suffer a  strong tidal interaction.
NGC~2992 is a disturbed Sa spiral seen nearly edge--on at an
 inclination of $70^\circ$. 
 On top of  having an active Seyfert~1.9 nucleus and a perturbed morphology due to the interaction, 
NGC~2992 shows a number of peculiarities. A prominent dust lane crosses most of 
the galaxy from the North--East to the South--West (see Fig.~\ref{fig:BVR}). It could be
 the remnant of a past merger. The \HI\ distribution is rather complex. On top of the
 expected rotating disk, fast--moving clouds are observed. Finally, the galaxy exhibits 
 several ionization filaments that are located well beyond the ionization cone 
associated with the AGN. They can be seen in Figure~\ref{fig:HaHI}. Their
kinematics is highly suggestive of outflows. Paper~II will be  devoted to
 the study of these phenomena which are likely more directly related to the nuclear activity
than to the interaction.

 NGC~2993, a face--on Sab, is situated at about 3\arcmin (27~kpc) to the South--East. On a true
color image (see Fig.~\ref{fig:BVR}), NGC~2993 appears much bluer than its companion. Its blue 
color index  (\BV = 0.3 mag) and its rather high far--infrared to blue
luminosity ratio (\Lfir/\Lb = 1.7) is suggestive of an active starburst.
The \HH\ content is however rather low \citep{Sofue93} and the far--infrared to \HH\ mass --- a measure of
the star formation efficiency --- is much higher than in classical starburst galaxies
\citep{Sanders85}. Active star--forming regions, as traced by \Ha , are present all over the
galaxy (see Fig.~\ref{fig:N2993:VHa}), but are more concentrated in the clumpy inner regions where
the \HI\ peaks as well. Some, red unresolved stellar clumps  can be seen in the outskirts of the
galaxy, especially to the North. They might be old globular clusters.

Our VLA map (Fig.~\ref{fig:VHI}) disclosed a third partner at roughly the same velocity,
  lying to the south--west which 
is listed in the Catalog of Flat Galaxies \citep{Karachentsev93} as
  FGC~0938. On our CCD image (Fig.~\ref{fig:F938:B}), it appears as an
 almost undisturbed edge--on disk with an ellipticity of 0.83, consistent
with an inclination of 80$\degr$. 
With  $\Mb=-15.5$,  FGC~0938 has the absolute blue magnitude of a dwarf
 galaxy. Its projected blue central surface brightness is 22.5~\sbu.
 Its rotating \HI\  disk looks regular apart from perhaps a
small extension to the West. The peak rotational velocity is of order
  60 \kms, corroborating its classification as a dwarf galaxy. Despite
their different environments, 
FGC~0938 has many properties in common with the  superthin galaxy UGC~7321   
recently studied by \citet{Matthews99}. It would be remarkable if a galaxy involved
in a tidal interaction would manage to keep a flat disk.
 This is a strong indication that FGC~0938 is presently
heading for its first encounter, falling towards the NGC~2992/93 system.

\subsubsection{The tidal features}
Arp~245 features three major tidal structures resulting from the interaction between NGC~2992 and NGC~2993:
two long tidal tails escaping from both spiral galaxies and a bridge between them. Their properties 
are summarized in Table~\ref{tab:tidal}.
 The northern  tail
 associated with NGC~2992 is about 16~kpc long, a rather modest extent compared to the 100--kpc long antennae
 in the prototypical 
interacting systems NGC~4038/39 and NGC~7252. This might be due to the smaller size and earlier  morphological
type of the progenitor and to the fact that  the interaction is witnessed in its early phase when the 
tails did not have time to fully develop.
 The northern tail has roughly the same distribution and extent in
both its stellar (optical) and
 gaseous (\HI) components. The optical tail looks like a plume, spreading at its tip where the \HI\ peaks
at a column density of $2 \x 10^{21}~\cmm$. 
This sub--structure, detected in all optical and near--infrared bands,
suggests the formation of a tidal dwarf galaxy. It hosts
several star--forming regions, clearly visible in the \Ha\ map (see Fig.~\ref{fig:TDG:Ha_spec}) at
 the location of the  \HI\ clump. The integrated colors of the tail as
a whole (\BV = 0.57, \VK = 0.42) are similar
 to the colors of the
outer regions of NGC~2992 (\BV = 0.62, \VK = 0.43). Some slightly bluer colors by 0.1~mag are
 measured within the \HII\ regions. The stellar population of the tail is hence 
 dominated by old stars pulled out from the parent galaxy with some weak contribution from young
stars born in situ in the tidal object.
 The TDG candidate will be studied in more detail in Sect.~\ref{sect:TDG}. 

A diffuse stellar bridge, with a maximum V surface brightness of 24~\sbu, connects the two
 spiral galaxies. The orientation of its \HI\ counterpart seems to be slightly twisted.
 No stellar clusters no
\HII\  regions or \HI\ substructures are found there. Instances of star--forming bridges do exist 
 \cite[e.g., NGC~6845; ][]{Rodrigues99}  but are in general quite rare among interacting systems.

The tail emanating from NGC~2993 to the East is highly curved in the plane of the sky.
 This is particularly visible in its \HI\
component which shapes like a ring. The projected total extent of the NGC~2993 tail is 30~kpc.
The  optical tail appears  sharp edged at its base and more diffuse further out.
The southern and western parts of the \HI\ ring--like structure have no optical counterparts up
to a limit of \sbV=25~\sbu. No obvious \HII\ regions are found along the tail, at 
least over the area covered by our \Ha\ image. Further out,
 the absence of bright optical condensations in the broad band images strongly
suggests that star formation, if any, is very weak there,  contrary to the NGC~2992 tail.
 The main difference between the
two tidal features is their respective \HI\ column density which turns out to be much lower in the
 NGC~2993 tail. The few \HI\ condensations scattered  over the ring have all peak column densities
lower than $3 \x 10^{20}~\cmm$. Surprisingly, the gobal color of this
quiescient tidal feature ($\BV = 0.39$) is    bluer than that of 
the northern star--forming tail and similar to that of the outskirts of its parent
galaxy, NGC~2993, a  starburst spiral. The  blue color of the tidal
tail might therefore reflect the past star--forming activity in the parent galaxy.

Our VLA map revealed one more feature with, perhaps, a tidal origin
 and lying east of the NGC~2993 ring, 
 a detached \HI\ cloud
with a mass of $6.5 \x 10^{7}~\Mo$. This object, visible in several
 VLA channels, and therefore likely real, has no
optical counterpart.

In total the \HI\ in tidal features contributes to about 45\% of the total \HI\ mass measured
in Arp~245. The \HI\ mass of the northern tail is about 60\% of that of
 NGC~2992\footnote{The mass of the HI clouds seen in absorption towards the nucleus of
NGC~2992 has not been included} whereas
the  \HI\ mass of the southern ring is 80\% of that of NGC~2993.  
Such  large values are typical of
gas--rich interacting systems where the contribution of the tails may
reach 90\%  
\citep[such as in Arp~105; ][]{Duc97b}. On the other hand the  tidal features contain a much 
weaker proportion of stars. The tail associated with NGC~2992 and NGC~2993 contains 
 respectively 17\% and 13\% of their stellar luminous population.
Such values have been determined from the V--band luminosities, assuming that the
M/L ratio is the same for all objects.

The 2D kinematics of the southern tail derived from the \HI\ datacube is quite smooth
 (see the channel map in Fig.~\ref{fig:HIchannel} and the velocity map in Fig.~\ref{fig:HIvel}).
The ring shows a coherent velocity field which is matched by our
 simulations (see below), as does the northern tail just south of the tidal dwarf.
 We will show in Sect.~\ref{sect:TDG:kinematics}
 that the kinematics of the TDG itself is probably decoupled from that of its host tail.

\subsubsection{Other unrelated objects}
\citet{Weedman71} reported the presence of a blue stellar object, Weedman~2 or
[BOB94] 0943-1403 \citep{Bowen94}, near the tip of the tidal
 plume of NGC~2992. Based on low resolution spectroscopic data, 
\citet{Burbidge72} claimed that it could be a QSO somehow ``associated'' with NGC~2992.
We obtained in January 2000 with EFOSC2 at the ESO 3.6m telescope a spectrum of this
object. It turns out to be a star.
During the same run, we measured  the redshifts of two uncatalogued background galaxies that
are situated North of NGC~2992 at resp. $\alpha = 09^{\rm h}45^{\rm m}45.55^{\rm s},
 \delta=-14^\circ16^\prime23.7^{\prime\prime}$ and
$\alpha = 09^{\rm h}45^{\rm m}51.14^{\rm s},
 \delta= -14^\circ15^\prime43.8^{\prime\prime}$ (J2000).  Both lie at a redshift of $z=0.111$.

\subsection{A numerical model of the interaction}
In this paper, we are mainly interested in a first, approximate model of the
NGC~2992/93 system, capturing its essential features while not
necessarily providing a perfect fit to all its details. In a future
paper, we plan to systematically refine this model and improve on its
ability to fit the finer features of the system.  The model developed here
is particularly useful to date the various phenomena observed in
Arp~245, to reconstruct its history and to predict its ultimate fate.

\subsubsection{Parameters of the collision}

The first encounter models based on restricted N--body simulations date 
from almost thirty years ago when \citet{Toomre72} were able to approximately match
the morphology of four well--studied interacting systems. 
Since then, numerical methods have become much more sophisticated but  the number of
interacting systems  with tidal tails that have been modeled in  a self--consistent way
 with numerical simulations is still small \citep[see e.g.,][]{Salo93,Thomasson93,Mihos93,Hibbard95b}.
  The main reason for that is the huge parameter space spanned by the numerous
free parameters of any such model (for instance, those defining the orbital plane),
 precluding simple methods to find a
matching solution for an observed system.
 Moreover the parameters describing
the structure of the colliding galaxies might not be well constrained by the
observations.  In that respect, Arp~245 appears as a particularly attractive
system to model. The morphology and dynamics of each of the interacting partners are
 indeed fairly well known.
 The large parameter space of possible collision
scenarios can thus be reduced drastically by making reasonable assumptions
and ``educated guesses'' about the colliding galaxies.

A sketch of the adopted geometry for our simulations is shown 
in Figure~\ref{fig:geometry}. Coordinates where the orbital plane
coincides with the xy--plane have been used while the orientations of the
spin vectors of the two disk galaxies have been specified in terms of
ordinary spherical coordinates $(\theta,\phi)$.

The nearly face--on orientation of the disk of NGC~2993, within
$20^\circ$, and the
orientation of its tidal arm emanating in the north--west suggest that
its spin vector points into the plane of the sky.  We have therefore
identified the plane of the sky with the plane of the disk of NGC~2993,
i.e., the observer is located in the opposite direction of the spin
vector of NGC~2993.
         The disk of NGC~2992 is seen about 70$^\circ$ inclined to the plane of
the sky with the upper east side being closer to the observer \citep{Chapman00}. Hence
we have assumed that the spin vectors of the disks of NGC~2992 and
NGC~2993 enclose exactly an angle of 70$^\circ$ with each other. 
       The prominence of both tidal tails suggests a largely
prograde encounter of the two galaxies, with the vector of the orbital
angular momentum presumably lying somewhere within the cone spanned by
the two spin vectors, or at least relatively near to it.  To further
simplify things, we have here assumed that the orbital angular
momentum lies in the plane spanned by the two spin vectors.

The velocity field shows that the center of NGC~2992 and  its northern tail
 is blue--shifted with respect to
NGC~2993. Also note that the well--developed tails of the two galaxies
indicate that the system is seen after its
first encounter (as will be confirmed below). In this phase, NGC~2992
 and its tail can be expected 
to move towards the negative y--axes. 
     The tidal response of the stellar disk of NGC~2993 seems
relatively weak compared to that of NGC~2992 which is in the process
of forming a dwarf galaxy out of its tidal debris.
 This suggests that NGC~2992 is
colliding almost fully prograde.
 Finally, we have
restricted ourselves to parabolic, zero--energy encounters.
The angular momentum of the orbit has been expressed in terms of the minimum
separation $b$ of a corresponding Keplerian collision.

 The galaxies themselves were modeled with a
massive and extended dark halo with an adiabatically modified
NFW--profile, an embedded exponential stellar disk, a stellar bulge,
and a gas distribution in the disk \citep[see details on the numerical
representations of these models  in ][]{Springel99}. Note
that in many observed disk galaxies the \HI\ gas is
substantially more extended than the exponentially distributed stars.
To be able to follow the gas at large radii, we have split the
available gas into two components of equal mass, one distributed just
like the stars and one forming a more extended component with a
constant surface mass density.

\subsubsection{Numerical simulations}

Using fully self--consistent simulations carried out with the Tree--SPH
code {\small GADGET}, we have coarsely explored the parameter space
remaining under the above assumptions  until a best--matching solution was found.
More specifically, the numerical model presented below is
specified by the following parameters.  The virial velocities of the
two galaxies were set to $V_{\rm c}=120\,{\rm km\,s^{-1}}$, i.e.~each
of them has a total mass of $4.0\times 10^{11} h^{-1}{\rm M}_\odot$.
We have assumed that a fraction $m_{\rm d}=0.05$ of the total mass resides
in the disk, 30\% of it in the form of gas, the rest as collisionless
stars. A fraction $m_{\rm b}=0.016$ of the total mass has been put into
a central stellar bulge.  Adopting a spin parameter\footnote{The spin
parameter $\lambda$ is defined as $ \lambda= \frac{{L |E|}^{1/2}}{G
M^{5/2}}$, where $L$ is the angular momentum, $E$ the total energy,
and $M$ the mass of the galaxy.} of $\lambda=0.06$,
the resulting exponential scale length of the disks is $R_{\rm
d}=2.7\,h^{-1}{\rm kpc}$. Half of the gas has been distributed in the disk
like the stars, and the other half in a disk of radius $8R_{\rm d}$
with constant surface mass density.  For the orbit of the collision,
we adopted the parameters $\theta_1=50^\circ$, $\phi_1=90^\circ$,
$\theta_2=20^\circ$, $\phi_2=270^\circ$, and $b=4 h^{-1}{\rm
kpc}$ (see Fig.~\ref{fig:geometry}). The numerical simulation shown
here used 60000 particles for 
the dark matter, 40000 for the stellar disk, 10000 for the stellar
bulge, and 40000 for the gas. Cooling of gas, star formation and supernova 
feedback were modeled as in \citet{Springel00}.  The cooling function used
 to describe radiative dissipation 
effectively cuts off at $10^4$~K, when the gas becomes neutral.
Some of the gas can reach somewhat lower temperatures by 
adiabatic expansion. Star formation is assumed to convert gas 
into collisionless stellar material at a rate given by a simple
Schmidt-type law, and supernovae feedback is modeled by an
effective turbulent pressure term. A full description of the
thermodynamic properties of this model is given in \citet{Springel00}.

The simulation begins when the galaxies' dark halos just start
to touch, i.e., at a separation of $240\,h^{-1}{\rm kpc}$. It then
takes about $1\,h^{-1}{\rm Gyr}$ to reach the perigalacticon for the first
time, when
strong tidal forces eject stars and gas out of the disks.  This
material forms the pronounced bridges and tails in Arp~245 when the
galaxies separate again. Eventually, they fall back together for a
second encounter, which takes place about $700\,h^{-1}{\rm Myr}$ after
the first collision. One hundred ${\rm Myr}$ later the galaxies completely coalesce
and form a single merger remnant. This time evolution of the system is
shown in Figure~\ref{fig:evolution}.

From Figure~\ref{fig:evolution}, it is clear that we observe the
system NGC~2992/93 at a time between the first and second encounter.
In this phase of the collision, the tidal features are very well
defined, whereas they are going to be much more diffuse at later stages of
the collision, which is a general result from simulations of other
systems. Our numerical
simulation provides the best match for the morphology
of the system at a time of around $100\,h^{-1}{\rm Myr}$ after the first encounter,
i.e. $1.1\,h^{-1}{\rm Gyr}$ after the start of the
simulation\footnote{Near perigalacticon the timestep in our adaptive  timestep
code is of order $0.05\,h^{-1}{\rm Myr}$.}. Figure~\ref{fig:sim:loc} shows
the face--on projection of the numerical model at this time.
After matching the morphology it becomes of course interesting to study
the velocity field of the gas 
distribution at this moment of the interaction.  As Figure~\ref{fig:HIvel}
shows, the overall pattern of gas flow is well reproduced by our
model.  Also, the striking ring of gas around NGC~2993 has nicely
formed in our model. Most of this gas is simply stemming from the long
tail which is pulled from the extended H{\small I} disk. The
inclination of the disk relative to the orbital plane helps to curl up
the tail in projection to an almost closed ring. There is actually
some shocked gas from the region of the bridge that helps closing the
ring.  

\subsubsection{Limitations of the model}

The simulation shown in this
paper corresponds to the model which so far provided the best match to
the morphology of the encounter.  While we are confident that the
orbital geometry is reasonably well determined in this model, the
internal structure of the galaxies is less well defined and may be
subject to revision when we refine the model using more detailed
comparisons between the simulation and the observational data.
In particular, we have here assumed that the two colliding galaxies
have the same mass and the same internal structure, an assumption that
is unlikely to hold in reality.

Besides, the dwarf galaxy FGC~0938 has not been taken into account. 
Having preserved a flat disk, this galaxy is most probably plunging
into the system and has not yet faced its first encounter with NGC~2992/93.
 Therefore, this intruder has not yet affected  the interaction
between the two spirals much. Later on, this galaxy might however slightly perturb
the merging history of the system.

Perhaps the main shortcoming of the present model is that it does not
form the tidal dwarf galaxy seen in the northern tail of NGC~2992 
although there is at least some gaseous overdensity at about the right
place. It remains to be seen whether the tidal dwarf can be produced
by a simple encounter model like the one examined here or whether
additional physics or more sophisticated collision scenarios have to
be invoked.

\subsection{The tidal dwarf galaxy candidate, A245N}
\label{sect:TDG}
 In this section we will study in detail 
the properties of the tidal object observed at the tip of the northern
tail of Arp~245. We will henceforth refer to it as A245N.

\subsubsection{Morphology and structural parameters}
\label{sect:morph}
Identifying a tidal dwarf in its host tail is obviously a difficult task. Such a problem
which raises the basic question  of the definition of a tidal dwarf galaxy will be addressed
 in Sect.~\ref{sect:TDG:def}.
In a first approach, we have isolated the TDG  candidate based on a morphological criterium.
 We have considered as belonging
to A245N all stellar and gaseous material in the tail located inside the isophote \sbB=24.5~\sbu,
 north of  $\delta=-14\degr18\arcmin$ (see Fig.~\ref{fig:VHI}).
 This region contains the bulk of the atomic and ionized gas and appears to be ``detached'' in
optical images (see Fig.~\ref{fig:BVR}).
Table~\ref{tab:TDG} lists the main properties of A245N.
 Figure~\ref{fig:TDG:BRJVK} presents
 images of A245N at different wavelengths from the optical
to the near--infrared. The tidal object has been detected in all
 BVRJHK$^\prime$ bands.   With an absolute blue 
magnitude of $\Mb = -17.2$,  A245N belongs to the bright end of the dwarf galaxy population.
It is actually as luminous and as extended as the LMC. 
 Its surface brightness 
 profile, shown in Figure~\ref{fig:TDG:sbr}, has been computed in the B and R bands following
 \citet{Papaderos96b,Papaderos96}. It  is exponential up to a radial distance of 
25$\arcsec$ (3.7 kpc) and drops beyond.
The extrapolated central surface brightness in the B band is 22.4~\sbu\  and the exponential scale length of the 
disk is 19\arcsec\ or 2.8 kpc.
 Put on the archetypal absolute magnitude vs.  surface brightness and scale length diagrams,
 A245N occupies the locus of low surface brightness dwarf irregular galaxies 
\citep[see Fig.~9 and Fig.~10 in ][]{Patterson96}.

\subsubsection{Stellar populations}
 Tidal tails contain two basic types of stellar populations. The first category
includes stars  older than the age of the interaction. They were originally formed in the
parent galaxy from which they have been pulled out. Numerical simulations indicate that
 the stars now found at the tip of the tidal tails, i.e., in the TDGs, initially belonged
 to the outskirts of the parent disk. The second category is made of stars younger than
 the interaction formed in situ in the tails after the collapse of tidally expelled
\HI\ material. 
For nearby systems, deep color--magnitude diagrams would be the ideal tool to disentangle
 the first and second generation stars. At the distance of Arp~245, the galaxy
can unfortunately not be resolved into stars. One has to rely on techniques based on the 
comparison of integrated broad--band photometric data with predictions of evolutionary
synthesis models. 

As a first step towards determining the age of the populations of A245N, we have compared its
optical color with that of its parent galaxy. The B--R  profile of both objects are 
displayed together in Figure~\ref{fig:TDG:col}. The external region of NGC~2992,
 beyound $r=30\arcsec$ (4.5 kpc) appears to have the same color as the tidal object except in
 its inner 10\arcsec.
Their  spectral energy distributions remain similar over a larger wavelength range,
as shown in Figure~\ref{fig:TDG:sed} which presents the SEDs of A245N and of the outskirts of NGC~2992
 ($r>30\arcsec$).
 Each color index differs by less than 0.1~mag, within the photometric errors.
 Note that the data
have not been corrected for internal extinction as this is highly uncertain
(see Sect.~\ref{sect:SF}).
Hence, it is clear that  the stellar population of  A245N is currently
 dominated by  stars pulled out from the  disk of the parent galaxy.

When analyzed in detail however, the stellar population of A245N does not appear
to be completely uniform. First, the object hosts several stellar clusters showing a range of
 colors (see Fig.~\ref{fig:TDG:BRJVK}).   Their faintness and the large
 photometric errors prevent us from  deriving from this color spread differences in
ages or metallicities for the individual clusters\footnote{Obvious bright Galactic stars have been subtracted  
from the images. However, some faint foreground stars could
have been mistaken for  stellar clusters. The
brightest blue clump in the B--band image,  for which we could obtain a
deep optical spectrum, does not exhibit the strong emission
lines expected for an  \HII\ region. However the detection of faint absorption lines 
(in particular, Ca H\&K $\lambda\lambda$ 3924,3968) at the right velocity confirms that it is 
a genuine star cluster in the TDG.}. Moreover, our 
 spectroscopic data indicate that OB stars are present 
in A245N and therefore that the tidal object is currently forming stars.
Figure~\ref{fig:TDG:Ha_spec}a displays the \Ha\ map towards  A245N. Several individual clumps are
 concentrated in the northern region where the \HI\ emission peaks. Two of the five brightest
 have counterparts in the broad--band images (see Fig.~\ref{fig:TDG:BRJVK}) and
 could be associated with older star clusters.
 The current star formation episode weakly affects the global photometry of the galaxy. The \VK\ color 
 map shown in Figure~\ref{fig:TDG:BRJVK} (where the \Ha\ contours have been superimposed) indicates
 that  star--forming regions are bluer by about 0.1~mag with respect to quiescient regions where no
ionized gas is detected. This variation appears to be 
 very small compared to predictions of  photometric models. For a pure instantaneous starburst,
  the \VK\ index might change in the first 100~Myr by 1 to 3 mag, depending on the 
metallicity \citep{Leitherer95}. Taking into account the old stellar component,
 one would get however a smaller color evolution.

We have estimated quantitatively the relative contribution of young to ``old'' stars 
 using an evolutionary population synthesis model developed
at the G\"ottingen observatory \citep{Krueger91}. This code has been updated to include
 the bimodal star formation history of tidal dwarfs \citep[see details
 in][]{Weilbacher00}.
We have only considered  solar metallicities and a Scalo IMF.
We have first reproduced the spectral energy distribution (SED) of the underlying population of A245N, assuming
 that it is similar to the SED of the external regions of its parent galaxy, NGC~2992.
  The best fit, shown in Figure~\ref{fig:TDG:sed},  was obtained by a template 
 for an Sb galaxy  with a characteristic age of 5~Gyr. 
Note that this result, which is valid
 for the outer disk of the spiral,  is consistent with the global morphological type
 of the galaxy being classified as an Sa. The effect of extinction 
 is indicated in the figure.
 We have then  simulated an additional burst,
 varying its shape, strength and duration and tried to fit the SED of the TDG candidate.
Taking into account the photometric errors and the dispersion between the
 observed SED and the best model, we found that stars younger
 than  100~Myr, i.e., stars formed in situ in the tail, cannot contribute for more
 than 2\% to the overall stellar mass  of the tidal object that we estimated from our
model to be $3\x 10^{9}~\Mo$. 
 With such a low burst strength, A245N differs from most of the  tidal dwarf galaxies studied
so far. In particular, the southern TDG in the Arp~105 system appears much bluer than its
parent galaxy; young stars could amount in this object to about 25\% \citep{Fritze98} of the
stellar mass.
The very blue tidal dwarfs in the NGC~5291 system also seem to be completely dominated by
  stars formed in situ in an \HI\  tail \citep{Duc98b}. A possible
 explanation for the difference between those systems and Arp~245
 might be that the latter object
 suffered its first encounter only 100~Myr ago, which might not have
 been a long enough time span to go through multiple bursts of star
 formation in order for the young population to dominate the old one.

\subsubsection{Current star formation and extinction}
\label{sect:SF}

We have obtained longslit spectra of some of the \Ha\ condensations of A245N.
 Our two slit positions were roughly aligned along the knots labeled
 TDG~1 and TDG~4  (see Fig.~\ref{fig:TDG:Ha_spec}a). A third spectrum, of TDG~3, 
was obtained  during our MOS run at the ESO 3.6m. 
The spectra of TDG~3 and TDG~4 are shown in Figure~\ref{fig:TDG:Ha_spec}b and
 the spectrophotometric data corresponding to the  knots with the highest
signal to noise are listed in Table~\ref{tab:TDG:spec}. 
The detected emission lines have relative fluxes typical of \HII\ regions ionized by young OB stars.
 There is no doubt that the \Ha\ map towards the tidal dwarf traces
 regions of star--formation. 
Despite the strong contribution from  stars older than 100 Myr noted before, no absorption lines are
visible in the spectra. The low surface brightness of this stellar component
might account for this result.

A Balmer decrement  \Ha/\Hb\  of 5.7 $\pm$ 1 has been derived from the 
spectrum of TDG~4. It  implies an intrinsic  extinction of 
about $\Ab=2.6 \pm 0.6$ mag, or E(B--V)=0.6, a value much higher than the typical E(B--V)=0.1
of irregular galaxies \citep{Hunter82}
 and about twice the value towards 30~Doradus in the 
LMC \citep{Greve91}. The discrepancy could suggest a very localized
obscuration in the \HII\ regions and hence a clumpy dust distribution. 
Slightly higher absorptions have been derived in TDG~1 and TDG~3, albeit  with a
much larger uncertainty. Besides, 
\HII\ regions in spiral disks also show comparable values of E(B--V) \citep{Dufour80}.

  The  equivalent width of the \Hb\ line in TDG~4, about $12 \pm 2$~\AA, is consistent
 with a starburst age of 5--7 Myr, according to the models by \citet{Cervino94}
 for which an  instantaneous burst with no underlying old component and a solar
 metallicity have been assumed. 
Given the age of the interaction, these newly formed stars were born in
 situ in the tidal tail.

We have estimated the star formation rate of the TDG from the total line emission extracted 
from the narrow--band \Ha\ image. The fluxes have been decontaminated for \NIIt\ emission using
the mean \NIIt/\Ha\ line ratio measured along our long--slit spectra.
 The calibration of \citet{Kennicutt98a} has been used to derive the SFR from the \Ha\ luminosity
corrected for galactic extinction. It assumes a Salpeter IMF (0.1--100~\Mo) and solar
metallicity.
We obtain an estimate of 0.03 \Mo/yr. Taking into account
 the  intrinsic  extinction derived in the optical from the Balmer decrement, one would
 get a SFR of 0.13 \Mo/yr. 
The star formation rate per unit area (in kpc$^{-2}$), $\log(SFR/A)$, is -3.1 (-2.5 after
correction for extinction). 
The normalized SFR derived by \citet{Hunter97} for a
sample of Im galaxies covers a large range of 4 dex, between -5 and -1, with 
a relative peak at -3.5. The SFR we find for A245N is somewhat
higher than this and is comparable to that of the LMC/SMC \citep{Kennicutt86}.
But despite these similarities,  A245N appears much redder than
star--forming Irrs and
more typical of a spiral. The simple fact that the global photometry of the tidal dwarf does not seem 
to be much affected by the  episode of in situ star formation implies that the latter cannot have
 lasted long with a constant SFR of $\sim$ 0.1 \Mo/yr. Given the short time scale available ---
less than 100~Myr, a constraint set by the simulations --- it is likely that the starburst
started only late, perhaps less than 10~Myr ago, as suggested by the equivalent width
of the \Hb\ line.

\subsubsection{Gas content}
A245N is a gas--rich object. Its total \HI\ reservoir of $9 \x 10^{8}~\Mo$ could
 sustain star--formation episodes at rates of 0.1 \Mo/yr for several Gyr.
 Its \HI\ mass to blue luminosity of 0.7 is typical of dwarf irregular galaxies 
\citep[according to ] [ the median value for Sm and Im UGC galaxies is 0.66 ]{Roberts94}.
A245N is the first tidal dwarf where molecular gas has been unambiguously found. 
Extended CO(1--0) emission has been detected with the IRAM 30~meter antenna over a region 
of 40$\arcsec$ centered on the \HI\ peak \citep{Braine00}. The \HH\ mass derived from
the combined CO line flux is $1.4 \x 10^{8}~\Mo$. 
\citet{Braine00} argue that the molecular
gas has formed in situ in the \HI\ clouds, directly fueling the star--formation regions.
The very similar width of the CO and \HI\ lines supports this idea.

The star formation efficiency, i.e the ratio of the star formation rate to the available 
molecular gas, \LHa/\MHH\ is 0.006, a  value  much closer to that  of early type spirals
 (about 0.01 \LHa/\MHH) than to that of
 irregular galaxies \citep[$\sim$ 0.05 \LHa/\MHH, ][]{Rownd99}. \citet{Smith99} detected
 molecular gas in 
the extended tail of the interacting galaxy Arp~215 (NGC~2782) where they derived
 a star formation efficiency three times lower than in A245N. Whereas the
  molecular gas in Arp~215 was found at the base of the tail and has probably been stripped
  from the disk of the parent galaxy, the CO detection in the tail of
NGC~2992 is more directly related to   A245N.
We have carried out a complete CO mapping of Arp~245  with the 15m--SEST antenna at la 
Silla in November 1999. The only CO cloud detected outside the parent disks is associated with 
 A245N at the location of the IRAM source, strongly suggesting that
the CO found here is related to the increase in density due to self
gravity in the tidal object and subsequent gravitational collapse of the tidal
\HI\ material.

\subsubsection{Metallicity}
\label{sect:metals}
We have estimated the oxygen abundance of A245N using the spectrophotometric data of 
the \HII\ regions TDG~3 and TDG~4 listed in Table~\ref{tab:TDG:spec}. A direct
determination of the abundance is not possible as the  temperature
sensitive \OIIIc\ line is not detected. Semi--empirical 
 calibrations were applied instead and several diagnostic diagrams  tested.
We first used the approximate calibrations of \citet{Edmunds84} based on the
 $R23 = (\OII + \OIIIt)/\Hb $ line ratio and applied it to TDG~3.
The relation between $R23$ and O/H is ambiguous as a single value
 of $R23$ corresponds to two values of the oxygen abundance. However, 
according to \citet{McGaugh94b}, the \NIIt/\OIIt\ line ratio provides
a useful diagnostic for choosing between the lower and upper branch of
the $R23$--O/H relation. log(\NIIt/\OIIt) $>$ -1 favors the highest 
metallicities. For TDG~3, we measured log$(\NIIt/\OIIt)=-0.4$, hence we
selected the upper branch, and derived from $R23$ an oxygen abundance 
12+log(O/H) of 8.6 $\pm$ 0.2. For TDG~4, the \OIIt\ line is outside the instrumental
 wavelength range.  Our estimate for this \HII\ region is therefore based on the 
less reliable $\OIIIt/\Hb$ line ratio as also calibrated by \citet{Edmunds84}.
We derived 12+log(O/H) $=$ 8.7 $\pm$ 0.2.
However, these calibrations are  not unique and depend very much on 
the ionization parameter $U$ \citep[e.g.,][]{McGaugh91} and on the temperature of the
 thermal ionization source $T$. We therefore decided to run
 a photoionization code trying to fit all our data on TDG~3 in a consistent
way. For this we used {\sc cloudy} \citep{Ferland96}. Our constraints were  the 
\OIIt/\Hb, \OIIIt/\Hb\ and \NIIb/\Ha\ observed  line ratios. We let the ionization
 parameter and temperature vary; we assumed a hydrogen density of 100~\cmm\
 and selected  a metallicity of about one half of solar (12+log(O/H)=8.6).
With log(U)=-3.4 and T=56200~K, the code  accurately  reproduced the oxygen line
intensities. The \NIIb/\Ha\ line ratio could only be obtained by  
 increasing the relative nitrogen abundance  log(N/O) to -1.1. This  abundance is quite  high 
compared to classical dIrrs and BCDGs \citep{Kobulnicky96} but consistent
with that measured in the outer regions of spiral disks  \citep{Ferguson98}.
 Models with solar metallicity failed to reproduce the \OIIt/\Hb\ line ratio.

 The spectrophotometric
 study of a sample of tidal dwarf galaxies  by  \citet{Duc99b} indicates that TDGs
 have an average oxygen abundance of \Zo/3 which is independent of their absolute blue magnitude
 (see Fig.~\ref{fig:TDG:met}).  Made of pre--enriched material, TDGs are more metal rich than
 isolated dwarf galaxies  of the same luminosity.
This property  may be used  to identify  recycled dwarf galaxies  and to investigate the
 origin of their building material in the disk of their progenitors, as discussed
in Sect.~\ref{sect:metenv}.

\subsubsection{Kinematics}
\label{sect:TDG:kinematics}
At first sight, the mean \HI\ velocity field  along the northern tidal tail
 (see Fig.~\ref{fig:HIvel}a)  appears to roughly follow the model field governed by
 streaming tidal motions (Fig.~\ref{fig:HIvel}b).
 Velocities range between 2270~\kms\ at the base of the tail
 to 2140~\kms\ at its tip.
 A detailed analysis of the \HI\ datacube
  shows however  a  possible evidence for a sub--structure associated with A245N
 which is visible in position--velocity
(PV) diagrams. This is best seen after rotating the HI datacube
 ($\alpha, \delta, {\rm V_{Hel}}$)
 clockwise by 15 degrees so that the tidal tail points upwards.
 Figure~\ref{fig:PV:TDG} presents three PV--diagrams taken along a
 direction perpendicular to the 
 tail axis. One cut is taken through the optical center of A245N, the other
 two are offset along the tail by $45^{\prime\prime}$. The cut through
 the tidal object shows a velocity gradient of $\sim 65~\kms$, peak to peak
 extending over almost $50^{\prime\prime}$ spatially.
 This  might  suggest that part of the tidal \HI\ is kinematically decoupled
from its host tail showing some evidence for solid body
 rotation with its kinematical major axis
 roughly perpendicular  to the tail. This component contains the bulk of the \HI\
 in the tail (more than 60\%) and corresponds spatially to the optical tidal object.
Due  the scarcity of \HII\ regions in
the tail, we could not determine,  based on our longslit data, any  velocity profile in the ionized gas
component.

   Based on our low resolution  \HI\ data only we estimate
a lower limit for its dynamical -- virial -- mass (i.e. uncorrected for
inclination)  of $9 \times 10^8\Mo$, identical within the
errors to the \HI\ mass and three times smaller than the estimated stellar
mass. This could indicate that a large fraction of the old stellar population present 
 in A245N does not belong to the  kinematically decoupled region of the galaxy. 
If reliable, our low value of the dynamical mass would be consistent with a very
low dark matter content, as expected for TDGs \citep{Barnes92}.

The \HI\ velocity dispersion towards the TDG candidate, 25--30~\kms, is much
 higher than in a quiescent disk.  The turbulence is hence dynamically important.

\subsubsection{Conclusions on the properties of   A245N}
Our detailed study   of A245N indicates that the tidal object has  apparently
 properties ranging between those of dwarf irregular galaxies 
 (structural parameters, gas content, star formation rate)
 and those of spiral disks (metallicity, extinction,  star formation efficiency, stellar population).
A straightforward explanation would be the intrinsic hybrid nature of the galaxy which
is  made of disk material but results from the collapse of gaseous clouds that
have masses and characteristics of dwarf galaxies.

\section{Discussion}
Tidal dwarf galaxies  candidates  have now been
found in a variety of interacting galaxies, i.e., in disk--disk systems
 \citep[e.g., NGC~4038/39, ][]{Mirabel92}, disk--spheroid systems
 \citep[e.g., Arp~105, ][]{Duc94a}, gas--rich spheroid--spheroid systems
 \citep[e.g., NGC~5291, ][]{Duc98b} and advanced 
mergers \citep[e.g., NGC~7252, ][]{Hibbard94}. These interacting systems have  a variety
of environments: field (e.g., NGC~4038/39), compact groups
 \citep[e.g., the Hickson groups, ][]{Hunsberger96}
 and clusters  (e.g., Arp~105, NGC~5291). On the other hand, not all
interacting systems with gaseous tails form dwarf galaxies, as recently shown by
\citet{Hibbard99}. Together with the overall properties of the parent 
galaxies and history of the collision, the environment --- i.e., density of companion galaxies,  density
of the intergalactic medium and strength of the associated ram--pressure ---  might play a role in the
 formation and more significantly in the evolution and survival time of tidal dwarf galaxies.
The exact contribution of each process is not yet known. 
 In that respect, Arp~245 is a simple system  situated in a relatively poor environment.
It may hence give us some clues about the required  {\it minimum
conditions} for the creation of such objects. Incorporating our data on Arp~245N with data from the
 literature on similar such systems, we will in the following
review some of the outstanding questions regarding TDGs, starting with a proper definition.

\subsection{Identifying Tidal Dwarf Galaxies in tidal features}
\subsubsection{A working definition of a Tidal Dwarf Galaxy}
\label{sect:TDG:def}
Luminous, star--forming knots  are commonly found along the tidal features
  of interacting systems. Will all of them become independent from their parent galaxies and hence
 become {\it true} galaxies ? Given their particular environments, one may
  doubt whether 
 all of them will manage to survive for more than one Gyr. Therefore,
  in order to distinguish between such a short--lived object and a
  true  ``Tidal Dwarf Galaxy''  we need a stricter definition.
We propose as a working definition for a  ``Tidal Dwarf
  Galaxy'' {\it an object which is a  self--gravitating entity, formed
  out of the debris of a gravitational interaction}.

This restrictive definition ensures that TDGs are not simply agglomerated debris of  collisions
 but that they are active objects that have their own dynamics and a
 potential well that is strong enough
 to sustain themselves against internal or external disruption for at
 least 1~Gyr. Such kinematically decoupled tidal objects and  associated rotating gas clouds
 have been found in the interacting systems Arp~105  \citep{Duc97b} and
NGC~5291 \citep{Duc97b}.

\subsubsection{Possible mis-identification of TDGs}

 Under the definition given above, a TDG should be considered a separate entity  based
on dynamical rather than morphological criteria. Tidal objects may not  always look as
 ``detached'' or appear as contrasted entities in optical images, nor
should they always have very distinct colors if they have not yet formed a
substantial proportion of luminous young stars.
 The true optical morphology of the galaxy might hence be hidden by an old unbound stellar 
component pulled out from the disk of the parent galaxy.
Computing the integrated properties of a TDG, one should in principle only consider 
bound tidal material corresponding to the kinematically detached part of the tail. 
In practice, this is difficult since it would require two--dimensional high--resolution 
 velocity data. The integrated properties of a tidal object might hence well be
contaminated.

Moreover, projection  effects should be taken into account.
Tidal tails are generally curved  (see Fig.~\ref{fig:sim:loc}) and 
 seen edge-on,  they exhibit at their apparent tip superimposed material
 from the near and back side. The resulting projected structure might be as luminous as a dwarf
 galaxy and hence be mistaken as a single object. Even the streaming motions of
an expending tail could mimic at its projected bend the dynamical signature of a
rotating TDG \citep{Duc97b}.  We could see there a range of velocities due to 
the contribution of several velocity vectors at different angles. It is however 
expected that, if this is the case, the resulting  velocity field should be 
asymmetric. 

Besides, some TDGs
could, in principle,  be mistaken with dwarf galaxies preexisting 
 the collision, and the tail linking them to a parent galaxy  would
 then  be a bridge.

\subsubsection{Is A245N a tidal dwarf galaxy ?}
Several observational
facts are inconsistent with the hypothesis that A245N is a preexisting dwarf galaxy.
 First of all, its high metallicity is a clear sign of a recycled origin.
 Its colors are remarkably similar to those of  the  parent's outside disk.
 Finally our numerical simulations
 show that no third body is required to reproduce the morphology of the
 interacting system.

 The only massive tidal object in Arp~245 is  
  found  at the tip of an almost edge--on tail, a case for which projection effects might be important.
 However  the \HI, \Ha, and CO lines
  peak at the same location in A245N and  have similar velocities. 
The projection hypothesis would require that all these components are rather uniformly
distributed along the tails which is not seen in the face--on eastern tail, at least
for the \HI\ gas. Besides, as discussed in Sect.~\ref{sect:sfthreshold}, ionized gas
is only observed  above a critical \HI\ column density. 
 The \Ha/\HI\ coincidence is a strong indication that both phases of the gas are physically linked. 
 It is hence more likely that the various gaseous components towards A245N form, at our spatial and
 velocity resolution, a single entity.
The surface brightness profile of A245N, which is 
very well fit by an exponential profile (see Fig.~\ref{fig:TDG:sbr}), might suggest the 
presence of a stellar disk. However, the profile expected for a pure tidal tail
without bound objects in it is not yet  well known. In our low resolution simulations, the SBP at the 
tip of the numerical tail seems to diverge from an exponential profile 
(see Fig.~\ref{fig:num:sbr}). Note however that the SBP measured towards A245N corresponds
to that of the old stellar component pulled out from the disk of its parent galaxy. Given the young
age of the interaction  and the non--dissipative nature of stars,  it would be surprising that the stellar
 population  is already relaxed and bound with the gaseous entity.

The kinematical independence of A245N is yet difficult to assess.
\HI\ position--velocity diagrams  show a fairly localized symmetric velocity gradient
 reminiscent of solid--body rotation with an axis perpendicular to the tail 
 (see Sect.~\ref{sect:TDG:kinematics}). But this is at the limits
of the spatial and velocity resolution of our VLA observations.
Besides, position--velocity diagrams of the simulated system computed at the same locations 
as the \HI\ ones  appear strikingly similar (see Fig.~\ref{fig:PV:TDG} and Fig.~\ref{fig:num:PV}).
Therefore, either a kinematically decoupled tidal object has already formed in our simulations, 
or the  \HI\ PV--diagrams mostly reflect  streaming tidal motions.
Clearly  higher resolution numerical simulations and \HI\ observations will shed more light on this.
 On the other hand,
the high velocity dispersion measured in the \HI\ component of A245N indicates that turbulence is
   almost as important dynamically as the  putative rotation. 

Most probably, A245N is a tidal dwarf galaxy observed in the early phases of its formation.
Its  gas might just be becoming self--gravitating, overcoming turbulence and streaming tidal motions.
This would be consistent with the young age of the interaction.

\subsection{Origin of the TDG building material}
\label{sect:metenv}
TDGs consist of material that has been pulled out from parent 
galaxies. But where exactly do the tidally expelled stars and gas
clouds come from? Numerical simulations by \citet{Hibbard95b} show that
 particles found at the end of the simulated tidal tails where TDGs are usually observed 
 once belonged to the outermost regions of the parent disk. This is
confirmed by our numerical simulations of Arp~245 (see Fig.~\ref{fig:sim:loc}).
The gas particles towards the tip of the TDG were originally in the outer gas
disk, at a  radius of about eight times the exponential scale length of
the stellar disk.  The simulations show however that the inner parts of the
tail come from regions at smaller radii in the disk.

The prediction of the simulations can
be observationally checked by noting that the
 metallicity of TDGs reflects that of the  region where their building
material has come from and been pre--enriched. 
 Spiral galaxies show strong metallicity gradients, from above solar in
the core to one tenth of solar in the outer regions
 beyond the optical disk \citep{Ferguson98}. An oxygen abundance of one third solar,    
 the median metallicity of the ionized gas in TDGs (see Fig.~\ref{fig:TDG:met}), roughly
 corresponds to a radius of $R_{25}$. The slightly higher metallicity of TDG A245N suggests an even 
smaller galactic radius.

 Hence, it is improbable
that most TDGs are made up of material from much further out, unless a 
  strong local enrichment has occured or unless the parent galaxy had 
 an unusual metallicity distribution  prior to the collision.  
Enrichment of the ionized gas by a burst of star formation within a time scale of 100~Myr or
 less -- the maximum age of the TDG A245N as determined from numerical simulations -- is unlikely.  
Studies of classical metal poor dIIrs and BCDGs have indeed shown that enrichment of the ionized
  gas does not seem to be efficient in low mass galaxies even over longer periods.
Parent galaxies with perturbed metallicity gradients, e.g. higher than
 normal, might be found in the class of Seyfert galaxies to which NGC~2992 belongs. Their active nucleus
 could  pollute the interstellar medium via large--scale outflows.
Whereas the absence of a steep abundance gradient has been reported by \citet{Evans87} in the 
 prototype Seyfert~2 galaxy NGC~1068, this property was not found
in other Seyfert galaxies \citep{Schmitt94,Storchi96}. 
  Statistical data are  so far missing to link any anomalous metallicity gradient distributions
 with the nuclear activity and related ejection phenomena. 

Clearly,  precise  measurements of the metallicity distribution in the parent galaxies
 \footnote{This study is extremely
difficult to do in NGC~2992 because of the absence of  \HII\ regions in its disk. Metallicity
measurements are very uncertain in the ELRs of active galaxies. } and
more detailed numerical simulations of interacting systems  would be required
to further investigate the origin of the building material of TDGs.

\subsection{Required conditions for the formation of a TDG}
\label{sect:cond:TDG}

\subsubsection{Location of TDGs in the tail}
Numerical simulations by \citet{Barnes92,Elmegreen93} suggest that TDGs should
form from gravitational instabilities that grow in the debris of the
collision. These objects appear as bound condensations distributed all
along the tidal features.
Some tails of interacting systems indeed host numerous faint blue
knots showing signs of star formation
\citep{Schombert90,Weilbacher00}. One of the two most spectacular objects of
that kind are NGC~4676, the Mice, \citep{Hibbard96,Sotnikova98} and
IRAS~19254--7245, the Superantennae \citep{Mirabel91,Mihos98}.
However, systems containing substructures that comply with our
definition of a TDG generally show the star formation being
concentrated in a single object located at the tip of the stellar
tail, such as TDG A105N \citep{Duc97b}. And this is also, of
course, the case for A245N, which is located at the end of the tidal
plume stemming from NGC~2992.

So, why is there only one single massive star forming clump as soon as a TDG
has been formed and why is it located at the tip of the tidal tail ?

 Excluding the projection effects that could lead to a mis--identification of a 
TDG at the tip of the tail,
part of the explanation   might have to do with the timescale for
tidal material to fall back to the merger. This rate scales as
$t^{-\frac{5}{3}}$ \citep{Hibbard95b} and is highest shortly
after perigalacticon. Therefore, bound objects that might have
formed at the base of the tail would have already fallen back towards
their progenitor, leaving as the only viable region for the formation
of a TDG the tips of tails.
Modeling the prototype merger NGC~7252 with N--body simulations,
\citet{Hibbard95b} have found that about half of the tidal material
situated at the base of the already formed tail (130 Myr after
periapse) falls back within 130 Myr.  Arp~245 is observed about
100~Myr after perigalacticon and only about 50 Myr after the tails
have developed (see Fig.\ref{fig:evolution}). Over a time scale of 50
Myr, about one fourth of the initial tidal material might have fallen
back already.  A more precise study of the temporal evolution of the
return of both gaseous and stellar material, based on our own
numerical simulations of Arp~245, will be studied elsewhere.

Another hypothesis could be that the initial clumps, proto--TDGs,
might have merged. Detailed high resolution simulations could test
whether the formation of TDGs proceeds  in a hierarchical way.
If true, short--lived tidal features would not offer a suitable
environment for the formation and survival of TDGs. In particular,
tails that have a life expectancy greater than 1~Gyr are more likely
to generate TDGs than bridges which only last for about $\sim
10^{8}$~yr \citep{Struck99}. Another obstacle faced by tidal
bridges, illustrated for instance by Arp~245, is their intrinsic low
surface density, a critical parameter as we will show in the next
section.

\subsubsection{Conditions for star formation}
\label{sect:sfthreshold}
The fact that all observed tidal dwarfs, i.e., kinematically
independent objects in tidal features, are star--forming objects is
partly a selection effect: TDG candidates have generally been selected
for their brightness and blue color.  On the other hand no bound tidal
gas cloud is yet known which is not associated with a stellar counterpart.
So the question raised in the previous section --- where do TDGs form?
--- could be addressed by answering another one: what conditions are
necessary for star formation (SF) to commence in tidal debris? 

\HI\ maps of interacting systems give some clues: all TDGs are found
in \HI\ tails and are roughly adjacent to local peaks in the \HI\
column density map. This fact is not specific to TDGs, of course, and
actually applies to SF regions in all types of galaxies, from spirals
to dwarf irregulars, and might simply reflect the fact that \HI\ is
the raw material for any SF event. However, not every \HI\ clump hosts
a SF region. The onset of star formation apparently depends on an \HI\
column density threshold, as first noted by \citet{Davies76} for the
LMC. Further studies of dwarf irregulars
\citep{Gallagher84,Skillman87b,Taylor94} and low--surface brightness
galaxies \citep{vanderHulst93,vanZee97} suggest that this threshold is
about 0.5 -- 1 $\x 10^{21}\cmm$.

 Neither the universality of this threshold, nor the underlying physics is
known. It would be surprising if it would be the same in spiral disks,
in dwarf galaxies, and in the more chaotic environment of interacting
galaxies. And a word of caution, when making comparisons, one should
aim for comparable linear resolution as the peak \HI\ column density,
because of beam dilution, decreases with decreasing resolution.

In the case of Arp~245 both tidal tails show a clumpy \HI\
structure. Star formation occurs only in the TDG above densities of
 2 $\x 10^{21}\cmm$. All condensations in the southern tail have
their peak below $5 \x 10^{20}~\cmm$ (see Fig.~\ref{fig:HaHI})
and no SF is associated with this tail. In turn, 
towards the spiral NGC~2993 ionized gas is observed above our
sensitivity limit of $0.4 \x 10^{-16}~\usbflux$ at those locations where the \HI\
density exceeds $10^{21}~\cmm$. The threshold hence appears to be two times
higher in the tidal dwarf than in the spiral disk.
 The connection between \Ha\ and \HI\
in the other spiral, NGC~2992, is more difficult to assess. Some
ionization filaments are observed at large galactic radii where the
\HI\ has column densities well below $5 \x 10^{20}~\cmm$.  However
this apparent decoupling between the two phases of the gas is mainly
due to the very nature of the ionizing source, probably hard UV
radiation from the central AGN or from a shock induced process
\citep[][Paper~II]{Allen99}. 

As mentioned, the physical origin for the empirical \HI\ threshold is
not well understood \citep[see review by ][]{Skillman96}. One possible
explanation is a gravitational instability, as elaborated by
\citet{Kennicutt89} in the case of spiral disks. Because of the
limited resolution and the obviously strong tidal forces, it is not
possible to test the validity of this explanation for TDGs.

Alternatively, the threshold in \NHI\ might correspond to a minimum
required column density 
of dust to shield the gas from the UV radiation field
\citep{Federman79,Skillman87b}. Given sufficient shielding, a molecular 
cloud can form from the \HI\, allowing star formation to proceed. This critical
density should be metallicity dependent since the dust--to--gas ratio varies with the
element abundance \citep{Franco86}. It is then expected that recycled tidal dwarfs
should have an   
   \NHI\ threshold lower than that of isolated metal--poor dwarf galaxies and similar to that of
 spiral disks. 
Table~\ref{tab:TDGs:HI} lists the \HI\  column density data available in the literature
for several star--forming TDGs. Column~1 shows the peak \NHI\ --- or a
range of values in case
several TDGs are present --- and column~2 the linear resolution. 
The majority of the TDGs so far studied have
a column density of 2 -- 3 $\x 10^{20}\cmm$, indeed three times lower than in
dIrrs \citep{Skillman87b,Taylor94}.
However, the  
metal rich, CO rich, and presumably dusty TDG belonging to Arp~245 has its \HII\ clustered
 in the central core of the \HI\ cloud where the \NHI\ exceeds $2 \x 10^{21}\cmm$.
Rather than stating that Arp~245N is the exception, one should perhaps
rather question the low star formation thresholds in the other TDGs,
noting that beam dilution for the other TDGs, which were observed at
much lower resolution, could partly account for this large
difference.

\subsection{Models and numerical simulations of TDGs}
Several theoretical models for the formation of TDGs have been proposed.
 \citet{Elmegreen93} suggest that massive \HI\ clouds form first in the outer disk of
 interacting galaxies.
 According to the Jeans criterion, the local 
high velocity dispersion induced by the collision allows the formation of clouds with masses
much higher than  in a quieter environment, up to $10^{8}~\Mo$. These clouds
 are then tidally expelled into the IGM where they  collapse. 
 \citet{Wallin90} put forward a geometrical torsion of the tail leading to
 a local density enhancement. \citet{Barnes92} propose a local  amplification of
pre--existing stochastic clumps in the  tidal tails that start accreting  low $\sigma$ material --- 
in particular  gas --- and become gravitationally bound. The collected material will then collapse
 until rotation takes over. 

Numerical simulations that might help selecting between these different theories
of the formation of TDGs should also take into account the observed properties
of these objects, and explain:

(a) knots, perhaps star forming, but not necessarily leading to TDGs,
along the tidal tails;
(b) star--formation whenever the  \HI\ clumps reach some critical
column density, the origin of which could either be a gravitational
instability within the tidally ejected gas, or shielding from the
UV--radiation field once clumps get dense enough;
(c) that in most systems eventually only one massive object per tail
forms, located at the end of it;
(d) the possible merging between individual preexisting clumps;
(e) the time scale for the formation of a TDG --- this can be quite
short: in Arp~245, a TDG developed in less than 100~Myr;
(f) the about $1/3$ of Solar metallicity in the TDGs;
(g) the formation of TDGs in pure \HI\ tails which is the case when
the interaction involves gas--rich early type galaxies from which
stars are much less efficiently pulled out than the gas \citep[e.g.,
in NGC~5291, ][]{Duc98b}.

\section{Conclusions}
We have presented a multi--wavelength study of the interacting system Arp~245
 (NGC~2992/93) and a preliminary numerical simulation of the collision. From 
 our observations and  models, we obtain  the following results:\\
a) The system is observed at an early stage of the interaction,
 $100\,{\rm Myr}$ after first perigalacticon and 700 Myr before the
final merger. At the current time two long tidal tails and a bridge have developed.
  N--body/SPH simulations reproduce fairly well both the optical and \HI\
 features, in particular the ring--like structure observed in the gaseous
component.
 The VLA map reveals a third member to the interacting system: an \HI\
rich dwarf galaxy, seen almost edge--on and  looking unperturbed. This
object is hence probably falling into the group.\\
b) \HI\ in emission is found all along the two optical tidal tails
where it accounts for respectively 60\%\ and 80\%\ of the \HI\ of
 their progenitors, 
 NGC~2992 and NGC~2993. The \HI\ shows a peak at the end of the NGC~2992
tail where a gas reservoir of $10^{9}~\Mo$ supplies a star--forming, tidal
 object, A245N. \HI\ is seen 
in absorption towards the active nucleus of NGC~2992.
 Star formation, as traced by \Ha,  occurs in the main body of
 NGC~2993. In the tidal features, it is restricted  to the
 tip of the northern tail at the location of the TDG A245N,
where the \HI\ column density is at least 1.5--2 $\x 10^{21}~\cmm$.\\
c) The global properties of  A245N range between those of dwarf irregular galaxies 
  as far as its structural parameters,
  gas content and star formation rate are concerned, and  those of spiral disks as for
its metallicity, extinction, star formation efficiency and stellar population. In particular,
 although the object is actively forming stars, and has a large atomic and
molecular gas reservoir, the bulk of its stellar population is still
 dominated by the old component pulled out from the parent galaxy. 
 Whereas  A245N appears to be a self--gravitating  entity, our data lack spatial
 resolution to probe its proper dynamics. Most probably, A245N is 
a tidal dwarf galaxy that is still in the process of formation.

\acknowledgements

We are grateful to the support astronomers and night assistants from the
2p2, NTT and 3p6 teams which have helped us
during our different observing runs at la Silla. Special thanks to 
Pierre Leisy for his precious support,  to Vassilis Charmandaris who did the
CO observations at SEST and to Polychronis Papaderos who has computed the
surface brightness profiles of our objects.
 This work has greatly benefited from discussions with
 Jonathan Braine, Uta Fritze--v.Alvensleben, Simon White and from 
 the very useful comments  of the referee, Curtis J. Struck.
P.--A. D. acknowledges support from the network Formation and Evolution of Galaxies
set up by he European Commission under contract ERB FMRX--CT96086 of its
TMR program. I.F. M. acknowledges support from CONICET/Argentina.
 This research has made use of the Lyon--Meudon Extragalactic 
Database (LEDA) supplied by the LEDA team at the CRAL--Observatoire de     
Lyon (France), as well as of the  NASA/IPAC Extragalactic Database (NED) 
which is operated by the Jet Propulsion Laboratory, California Institute 
of Technology, under contract with the National Aeronautics and Space  
Administration.


\clearpage

\begin{figure} 
\epsscale{0.58}
\figcaption[Duc.fig1.ps]
{Optical true color image of Arp~245. The image is a combination
of NTT + CFHT + ESO 3.6m B--band images (blue), NTT + CFHT + ESO 3.6m V--band images (green)
and  NTT + ESO 3.6m R--band images (red). Bright foreground stars have been masked. 
The field of view is 6\protect\arcmin \x 8.2\protect\arcmin (54 \x 74 kpc).
North is up and East to the left.
 The system consists of two interacting
spiral galaxies: NGC~2992 to the North and NGC~2993
to the South--East. The two galaxies, connected by a diffuse bridge, 
 exhibit each a long stellar tidal tail. The one escaping 
from NGC~2992 hosts at its tip a star--forming tidal dwarf galaxy candidate, A245N.
NGC~2993 appears bluer than NGC~2992 and is currently experiencing a
vigorous starburst. A prominent dust lane in the latter galaxy lane partly 
obscures  emission from the active Seyfert~1.9 nucleus. A blue edge--on dwarf galaxy,
 FGC~0938, is visible to the South-East.
\label{fig:BVR}}
\end{figure}

\begin{figure}
\epsscale{0.8}
\figcaption[Duc.fig2.ps]
{\protect\HI\ distribution in Arp~245. The VLA \protect\HI\ contours are superimposed
on a V--band image of the system. The contour levels are
1, 2, 3, 5, 10, 15, 20 $\x 10^{20}$ \cmm. The radio beam is indicated at the
top--left of the image. The optical background picture consists 
of recombined calibrated V--band images from different telescopes displayed
 with an intensity scale in \protect\sbu\ as indicated to the right.  
 Although there is an 
overall agreement between the distribution of the stellar
 and gaseous components, some differences can be noticed between the
two interacting galaxies.
Towards NGC~2992, the  \protect\HI\ shows a peak at the location of the tidal dwarf
candidate. \protect\HI\
is seen in absorption against the radio continuum from the nucleus. 
 Towards NGC~2993, the \protect\HI\ map has a ring--like structure. Its
western part has no optical counterpart. A detached cloud can
be seen east of the ring. The VLA observations revealed
at the same distance as Arp~245 a  gas--rich edge--on dwarf
 galaxy, FGC~0938 which is visible to the South--West.
This  object which looks remarkably unperturbed might be falling 
towards the NGC~2992/93 system.   
\label{fig:VHI}
}
\end{figure}

\begin{figure}
\epsscale{0.7}
\figcaption[Duc.fig3.ps]
{NTT H$\alpha$ + \protect\NIIt\ image of Arp~245. The \protect\HI\ map with the same contours
as in Fig.~\ref{fig:VHI} is superimposed. The colored background indicates the
 field of view covered by the narrow--band image. The \protect\Ha\ image is
displayed on a  
logarithmic intensity scale. The sensitivity limit is $0.4 \x 10^{-16}~\protect\usbflux$.
 Towards NGC~2993
and  A245N,  \protect\HII\ regions are concentrated within \protect\HI\ clumps and 
trace
star forming regions. The biconical structure of the \protect\Ha\ emission in the inner
 regions of NGC~2992 and extending to larger scales in the presence of numerous
ionization filaments betray  strong nuclear activity in this 
Seyfert--1.9 galaxy. 
\label{fig:HaHI}
}
\end{figure}

\begin{figure}
\epsscale{1.}
\figcaption[Duc.fig4.ps]
{Calibrated R--band image of NGC~2993. \protect\Ha\ contours are superimposed.
 Levels  differ by a multiplicative factor of 2.
The lowest contour is $0.4 \x 10^{-16}~\usbflux$. The interval between the tickmarks is 20\arcsec . 
North is up and East to the left.
\label{fig:N2993:VHa}
}
\end{figure}

\begin{figure}
\epsscale{1.}
\figcaption[Duc.fig5.ps]
{Calibrated B--band image of FGC~0938. Contour levels are separated 
by 1~\sbu. The lowest contour is 26~\sbu. The interval between the tickmarks is 20\arcsec . 
North is up and East to the left.
\label{fig:F938:B}
}
\end{figure}

\begin{figure}
\epsscale{1.}
\figcaption[Duc.fig6.ps]
{20--cm radio continuum map superimposed on the R--band image of Arp~245.
Levels are 1.8, 3.5, 6.8, 13.2, 25.7, 50.1, 97.3, and 189~\protect\mJyb\ .
\label{fig:rad}
}
\end{figure}

\begin{figure}
\epsscale{0.8}
\figcaption[Duc.fig7.ps]
{\protect\HI\ channel maps of Arp~245. The contours in each channel 
are superimposed on an optical DSS image of the system. Black contours
correspond to \protect\HI\ seen in emission and white contours
  \protect\HI\ seen in absorption.  The lowest level is $\pm$ 0.8 mJy~beam$^{-1}$
and the interval is 1.8 mJy~beam$^{-1}$.
The velocity of each channel in \protect\kms\ is indicated at the bottom
of each frame. The
VLA beam size is indicated in the upper left hand corner.
\label{fig:HIchannel}
}
\end{figure}

\begin{figure}
\epsscale{1.}
\figcaption[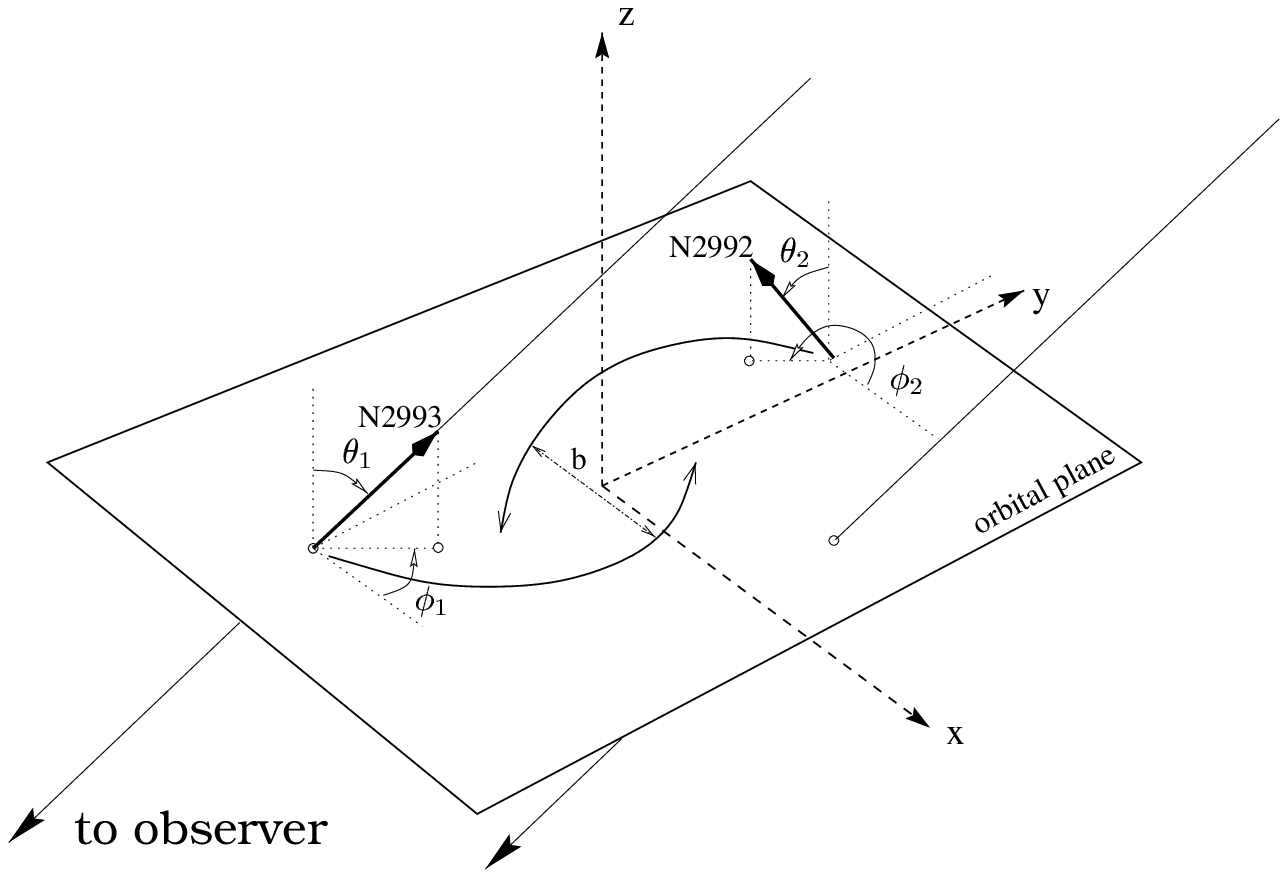]
{Sketch of the adopted orbital geometry for our simulations of the
system Arp~245. We set up the collision in such a way that the orbital angular
momentum is parallel to the z--axes. The initial galaxy positions are
chosen such that the galaxies would reach their minimum separation on
the x--axis, if they were point masses moving on Keplerian orbits.  The
orientations of the spin vectors of the two disks are specified in
terms of ordinary spherical coordinates $(\theta_1,\phi_1)$, and
$(\theta_2,\phi_2)$, respectively.  We shall assume that NGC~2993 is
seen exactly face--on and is rotating clock--wise; then the line of
sight is along the spin vector of NGC~2993.
\label{fig:geometry}
}
\end{figure}

\begin{figure}
\epsscale{0.8}
\figcaption[Duc.fig9.ps]
{Time evolution of our numerical model of the system Arp~245. In
each panel, we show the projected distribution of the stars
(color--scale) and that of the gas (green contours). Each panel is
$140\,h^{-1}{\rm kpc}$ on a side, and the labels give the elapsed time
after the start of the simulation in units of 0.1 Hubble times (or
$0.98\,h^{-1}{\rm Gyr}$). The model is shown in the plane of the sky and the 
orientation is the same as in Fig.~\ref{fig:VHI}.
\label{fig:evolution}
}
\end{figure}

\begin{figure}
\epsscale{1.1}
\figcaption[Duc.fig10.ps]
{Face--on projection of the numerical model of NGC~2992 at
 T=0.6 and T=1.1. The 5\% gas particles that 
are at the edge of the unperturbed disk  at T=0.6 are shown in red.
At T=1.1 they concentrate at the end of the tidal tail at the location of
 the TDG.
\label{fig:sim:loc}
}
\end{figure}

\begin{figure}
\epsscale{1.}
\figcaption[Duc.fig11.ps]
{Velocity map of Arp~245.
{\bf (a)} \protect\HI\ velocity map with optical contours superimposed. 
 {\bf (b)} Velocity field of our numerical simulation. The contours
indicate the distribution of the stellar component. The simulation is
shown at a time when it best matches the observed morphology and velocity field 
 of Arp~245. This point of time is $100\,h^{-1}{\rm Myr}$
after first passage through pericenter, and $1.1\,h^{-1}{\rm Gyr}$
after the start of the simulation. The color scale gives 
the line--of--sight velocity of the gas
relative to the systemic velocity of the system.
\label{fig:HIvel}
}
\end{figure}

\begin{figure}
\epsscale{0.65}
\figcaption[Duc.fig12.ps]
{Smoothed  calibrated BRJ images and V-K$^\prime$ color map of   A245N.
The surface brightness ranges in \sbu\ are indicated at the bottom
of each  image. Black contours are separated by 0.5~\sbu.
The \Ha\ contours are  superimposed in white.  Levels are 10, 30, 50, 70, 90\%
 of the peak flux ($4.0 \x 10^{-16}~\usbflux$). Tickmarks are separated by 10\arcsec. 
Some stellar clumps associated with \HII\ regions are clearly visible.
Foreground stars have been subtracted from the images.
\label{fig:TDG:BRJVK}
}
\end{figure}

\begin{figure}
\epsscale{1.}
\figcaption[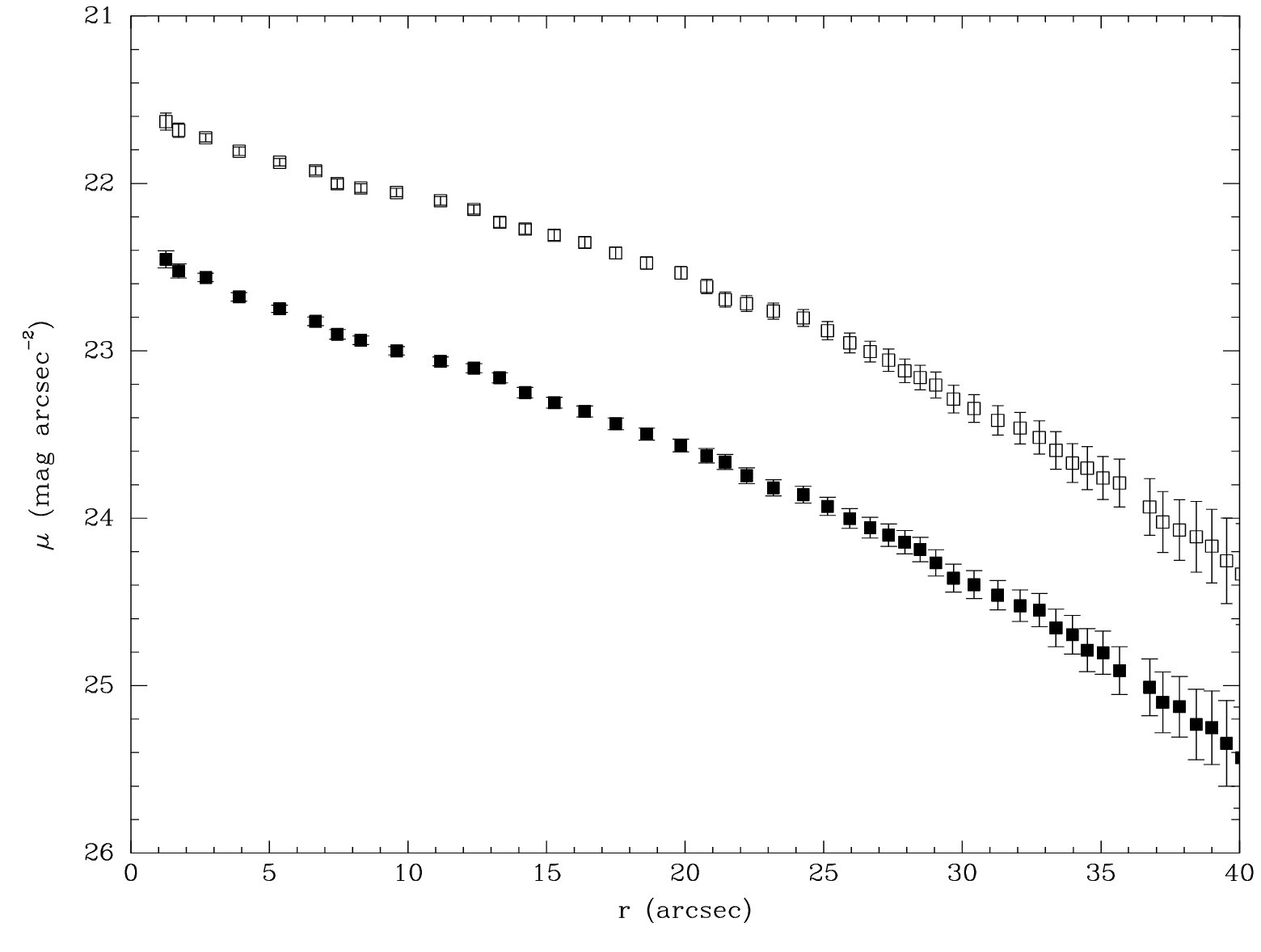]
{Surface brightness profile of  A245N in the B (filled squares) and
R (open squares) bands.
\label{fig:TDG:sbr}
}
\end{figure}

\begin{figure}
\epsscale{1.}
\figcaption[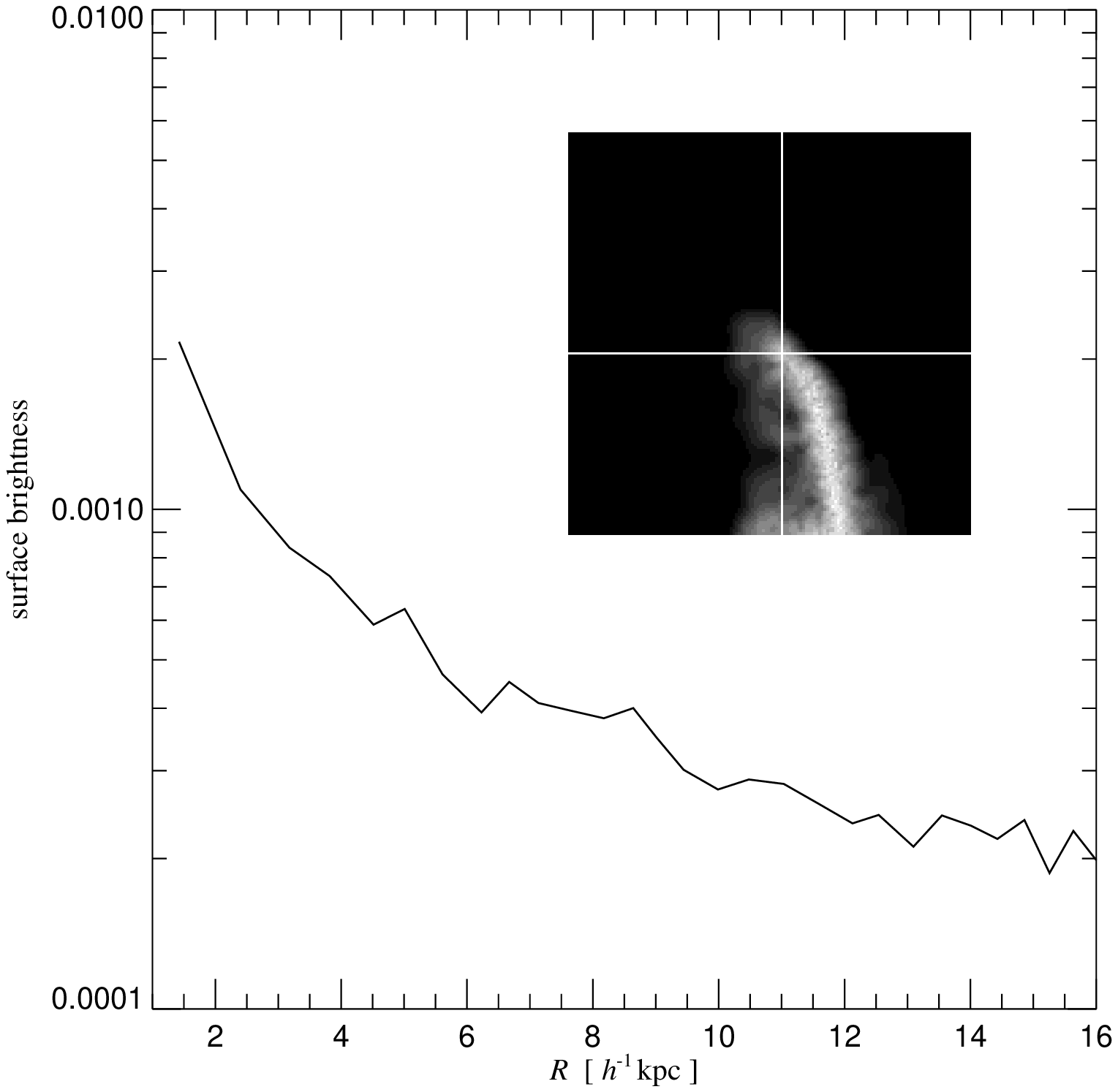]
{Azimuthally averaged surface brightness profile at the tip of the
simulated tail at T=1.1. The exact center  is indicated in the  surface brightness
 map.
\label{fig:num:sbr}
}
\end{figure}

\begin{figure}
\epsscale{1.}
\figcaption[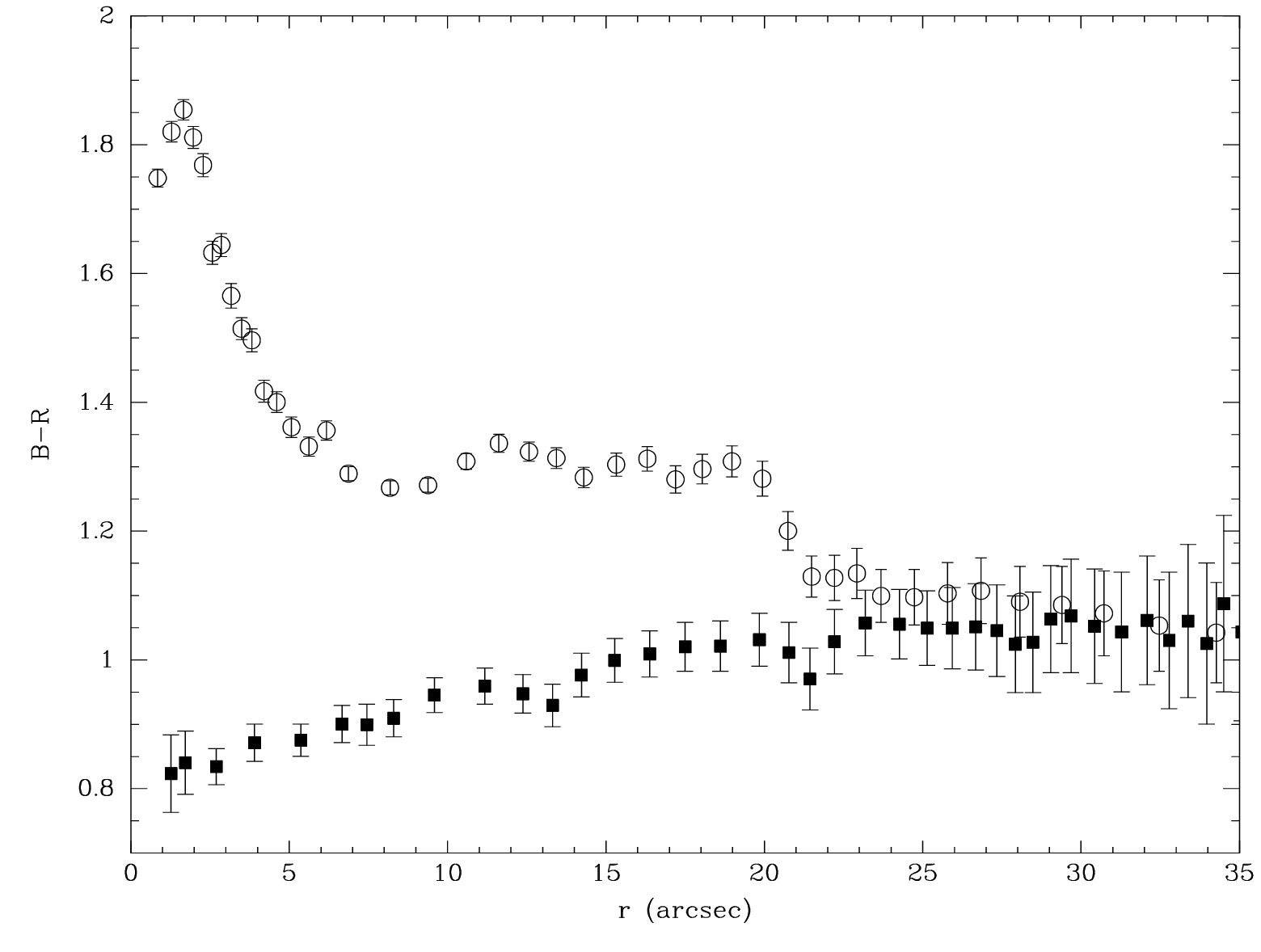]
{B--R color profile of  A245N (filled squares) and of its parent galaxy NGC~2992
 (open circles). The radius $r=0$ corresponds to the optical center of each object.
\label{fig:TDG:col}
}
\end{figure}

\begin{figure}
\epsscale{1.}
\figcaption[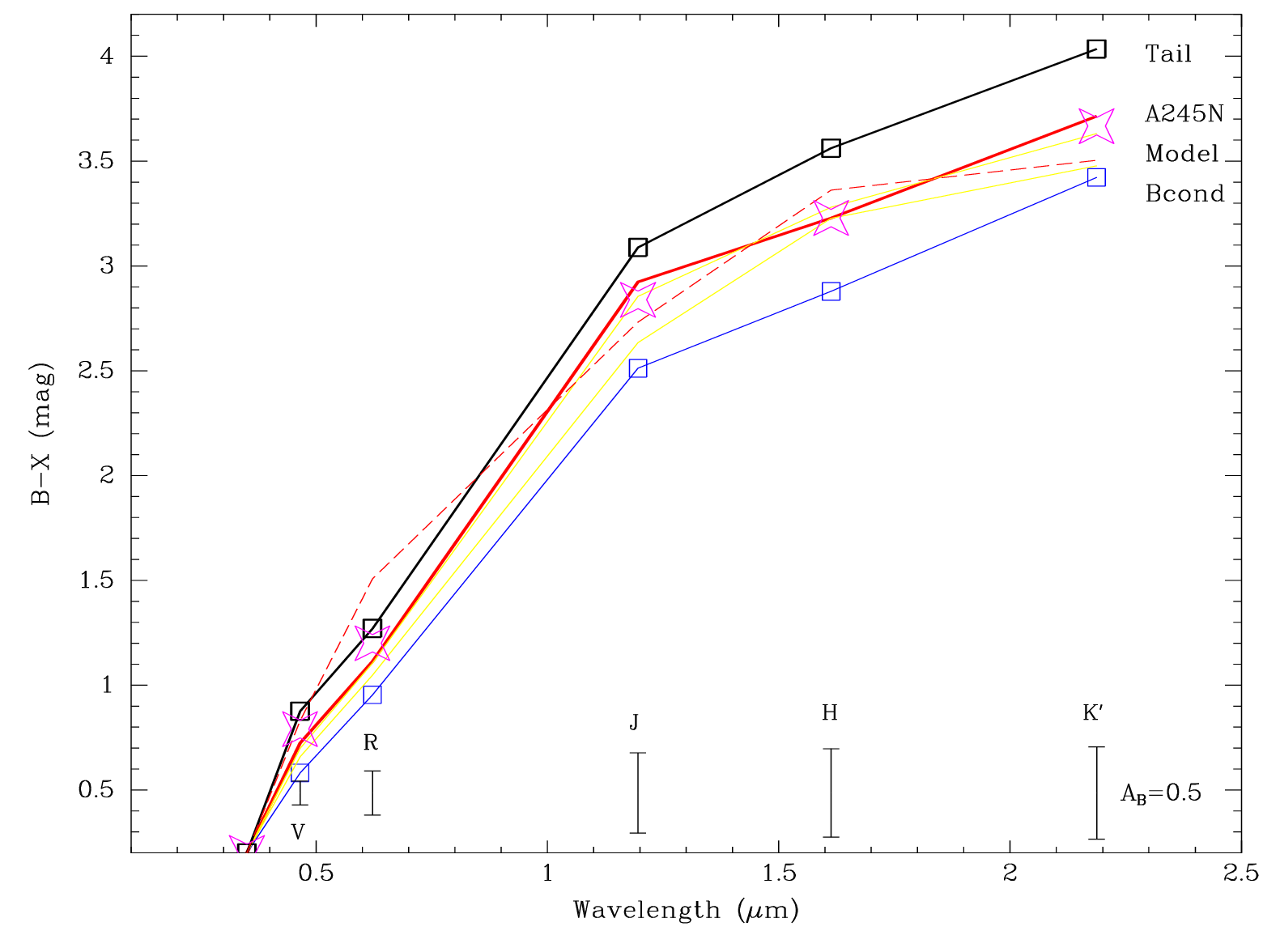]
{Spectral energy distribution of  A245N. The y--axis is the B-X color index
 in mag and the x--axis the wavelength in $\AA$. The SED averaged over
the entire TDG is
traced with the thick red line. The SEDs of the two brightest \protect\HII\ regions in A245N 
are indicated with the
thin yellow lines. The blue thin line corresponds to the SED of the bright blue stellar clump
 in A245N.
The SED of a region in the tail outside the \protect\HII\ regions is indicated with a 
thick black line. The purple stars correspond to the SED of the outskirts of
NGC~2992. Finally the red dashed line is our best model for the underlying stellar 
population of A245N. 
 The bars at the bottom correspond to an extinction of \Ab=0.5~mag. All 
magnitudes are corrected for Galactic extinction.
\label{fig:TDG:sed}
}
\end{figure}

\begin{figure}
\epsscale{1.}
\figcaption[Duc.fig17.ps]
{{\bf (a)} \protect\Ha\ map towards  A245N. The contour levels 
are 10, 30, 50, 70, 90\% of the peak flux 
($4.0 \x 10^{-16}~\usbflux$).
The principal condensations  are labeled. The interval between the tickmarks is 10\arcsec\ .
 {\bf (b)} Optical spectra of the clumps TDG~3 (arbitrary units) and
TDG~4 (the units are   \ufluxm\ ). The spectra, typical of  \protect\HII\
regions, show that the TDG is actively forming stars.
\label{fig:TDG:Ha_spec}
}
\end{figure}

\clearpage

\begin{figure}
\epsscale{1.}
\figcaption[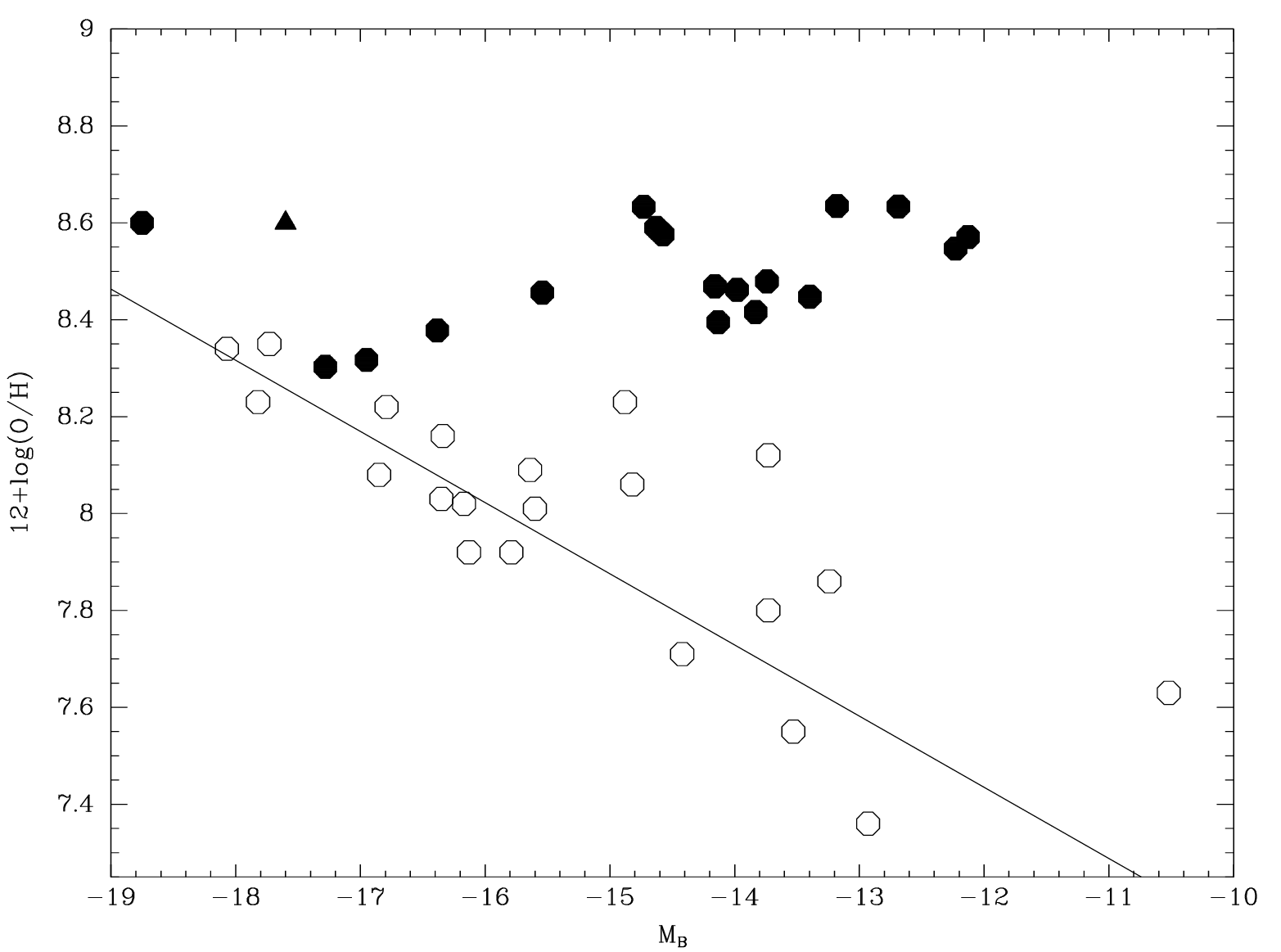]
{Oxygen abundance vs absolute blue magnitude for a sample of isolated 
nearby dwarf galaxies (open circles, Richer and McCall 1995) and 
tidal dwarf galaxies \citep[filled circles, ][]{Duc95}. A245N is indicated by
a triangle.
\label{fig:TDG:met}
}
\end{figure}

\begin{figure}
\epsscale{0.8}
\figcaption[Duc.fig19.ps]
{\protect\HI\ Position--Velocity diagram at three different positions
perpendicular to the northern tidal tail of NGC~2992, after rotating
the galaxy clockwise by $15^\circ$. The central panel goes through the
center of the TDG. The top and bottom ones are offset by
$45^{\prime\prime}$ to the north and south, respectively.
The lowest contour level is $\pm$ 0.5 mJy~beam$^{-1}$
and the interval is 2.3 mJy~beam$^{-1}$.
\label{fig:PV:TDG}
}
\end{figure}

\begin{figure}
\epsscale{1.}
\figcaption[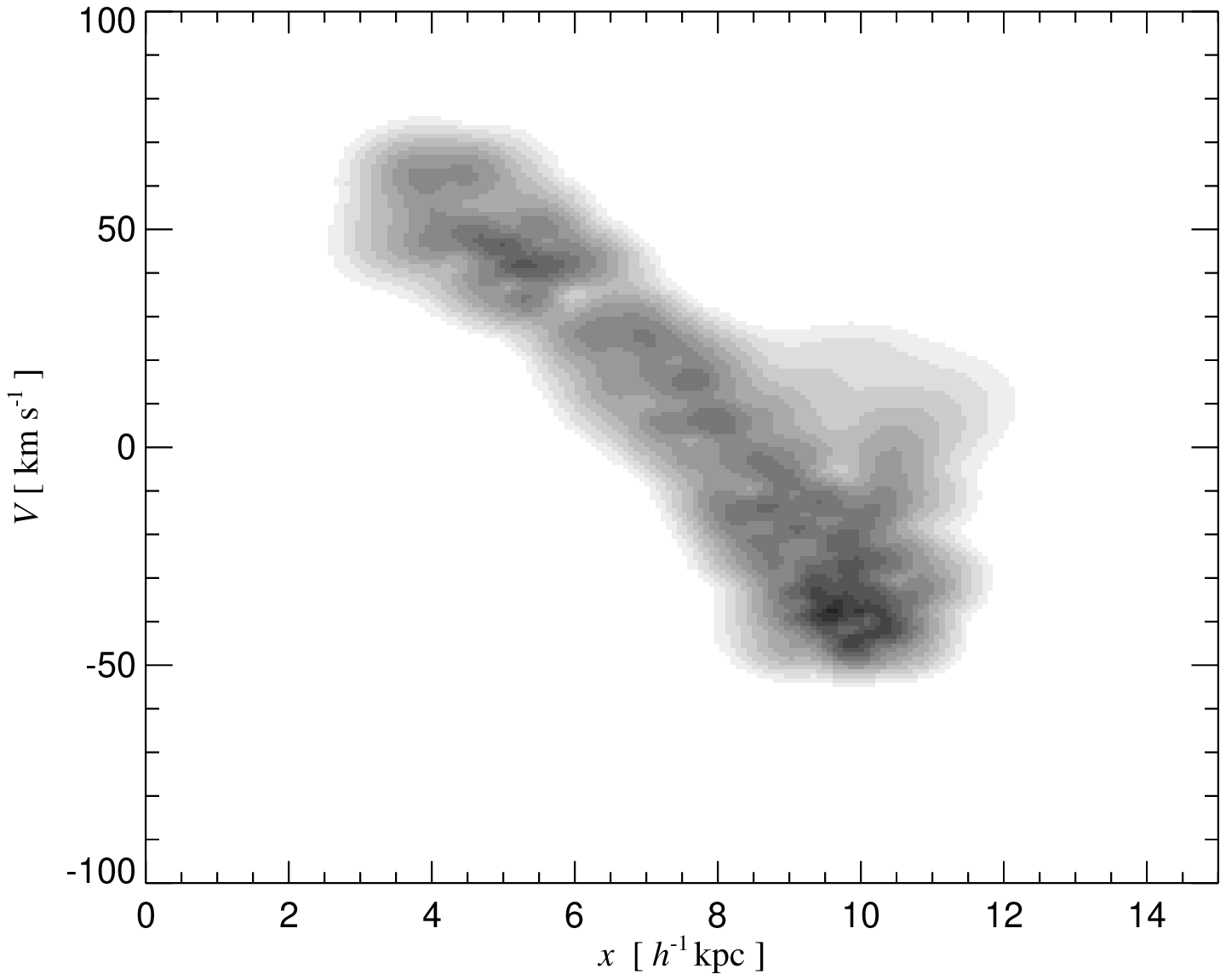]
{Simulated Position--Velocity diagram at a position perpendicular to  the northern tidal tail,
 after rotating the T=1.1 projected model clockwise by $15^\circ$.
The PV--diagram has been computed in  a thin slice of the gaseous component at $y = 32h^{-1}$ kpc.
\label{fig:num:PV}
}
\end{figure}

\clearpage

\begin{table}
\caption{Observing log.\label{tab:log}}
\begin{tabular}{llll}
\tableline\tableline \tabsp 
 & Date  & Telescope/Instrument & Technical details\\
\tabsp \tableline \tabsp
Optical imaging & Feb. 1992 & CFHT/PUMA  & \parbox{6cm}{1K SAIC1 CCD;
 0.34$^{\prime\prime}\,{\rm pixel}^{-1}$;\\ B,V filters}  \\
 \tabsp  \tabsp
                & Feb. 1995 & NTT/EMMI   & \multicolumn{1}{l}{\parbox{6cm}{2K Tektronix CCD;
 0.27$^{\prime\prime}\,{\rm pixel}^{-1}$; Bb,V,R filters (red arm)}}  \\ 
 \tabsp  \tabsp
                & Jan. 2000 & ESO 3.6m/EFOSC2 &  \multicolumn{1}{l}{\parbox{6cm}{2K Loral CCD;
 0.31$^{\prime\prime}\,{\rm pixel}^{-1}$ (2\x2 binning); B,V,R filters}} \\
\tabsp \tabsp
Near--infrared imaging & Feb. 1996 & MPG--ESO~2.2m/IRAC2B & \multicolumn{1}{l}{\parbox{6cm}{NICMOS3 IR array;
 0.5$^{\prime\prime}\,{\rm pixel}^{-1}$;\\ J,H,K$^{\prime}$ filters}}    \\ 
\tabsp \tabsp
H$\alpha$ imaging & Feb. 1995 & NTT/EMMI   & \multicolumn{1}{l}{\parbox{6cm}{2K Tektronix CCD;
 0.27$^{\prime\prime}\,{\rm pixel}^{-1}$;\\ HA/3,HA/0 filters}} \\  
\tabsp \tabsp
Long--slit spectroscopy & Feb. 1995 & NTT/EMMI   & \multicolumn{1}{l}{\parbox{6cm}{2K
 Tektronix CCD; 0.27$^{\prime\prime}\,{\rm pixel}^{-1}$ \\ 
 1.5$^{\prime\prime}$ wide longslit, grism \#3 (360 g\,mm$^{-1}$)}}\\ 
\tabsp \tabsp
MOS  spectroscopy & Jan. 2000 & ESO 3.6m/EFOSC2 & \multicolumn{1}{l}{\parbox{6cm}{1.7$^{\prime\prime}$ wide
 slits; grism \#11 (300 g\,mm$^{-1}$)}}\\
\tabsp \tabsp
\HI\ line map & Sept. 1997 & VLA/CS--array & details in Table~\ref{tab:VLA} \\ 
\tabsp \tabsp
CO line observations & Jun. 1999 & IRAM~30m &  \multicolumn{1}{l}{\parbox{6cm}{CO(1-0) beam: 22$^{\prime\prime}$\\
 details in Braine et al. (2000)}} \\ 
\tabsp \tabsp
                     & Nov. 1999 & SEST     &  \multicolumn{1}{l}{\parbox{6cm}{CO(1-0) beam: 44$^{\prime\prime}$\\ 
 details in Paper~II}}\\ 
\tabsp \tableline 
\end{tabular}
\end{table}

\clearpage

\begin{table}
\caption{VLA Observing Parameters.\label{tab:VLA}}
\begin{tabular}{ll}
\tableline\tableline \tabsp
Parameter &  Value   \\
\tabsp \tableline \tabsp
Object  & NGC~2992/93 (Arp\,245)   \\
Instrument  & Very Large Array (VLA)    \\
Configuration  & CS--array    \\
Observation date  & 1997 September 7    \\
Total observing time  & 5.7 hours     \\
Number of antennas  & 27     \\
Field center $(\alpha, \delta)_{J2000}$ & $09^{\rm h} 45^{\rm m} 42^{\rm s}, -14^\circ 19^\prime 35^{\prime\prime}$ \\ 
Flux/bandpass calibrator  & 1331+305  \\
Phase calibrator  &  0902--142   \\
Central velocity (V$_{\rm heliocentric})$ & 2341 km ${\rm s}^{-1}$    \\
Observed baselines (min--max) & 0.05 -- 3.4 km   \\
FWHP of primary beam & $32^\prime$    \\
System temperature & 35 K    \\
Correlator mode & 4    \\
Total bandwidth per IF &3.125 Mhz \ \ \  (650 km ${\rm s}^{-1}$)    \\
Number of channels per IF & 32    \\
Channel spacing & 21 km ${\rm s}^{-1}$   \\
FWHM velocity resolution & 21 km ${\rm s}^{-1}$    \\
RMS noise per channel & 0.4 mJy ${\rm beam}^{-1}$    \\
Smoothing applied & None \hskip 1.8cm Gaussian    \\
FWHP of synthesized beam & $19^{\prime\prime} \times 14^{\prime\prime}$ \hskip 1cm  $35^{\prime\prime} \times 35^{\prime\prime}$   \\
Conversion factor &  2.18 \hskip 2cm 0.495    \\
${\rm T_B(K)/S(mJy\,beam^{-1})}$ &     \\
Conversion factor & $3.97\times 10^{18} \hskip .9cm  9.02\times 10^{17}$ \\
 $N_{HI}({\rm cm}^{-2})$ $/N_{HI}({{\rm mJy~ beam^{-1}}}~ {{\rm \kms}})$ &     \\
\tabsp \tableline \tabsp
\end{tabular}
\end{table}

\clearpage

\begin{table}
\caption{Properties of the interacting partners.\label{tab:partners}}
\begin{tabular}{lccc}
\tableline\tableline \tabsp 
                                 & NGC~2992    & NGC~2993     & FGC~0938         \\
\tabsp \tableline \tabsp 
$\alpha$ (J2000)                 & $09^{\rm h} 45^{\rm m} 42.0^{\rm s}$ & $09^{\rm h} 45^{\rm m} 48.3^{\rm s}$ & $09^{\rm h} 45^{\rm m} 34.4^{\rm s}$  \\
$\delta$ (J2000)                 &$ -14^\circ 19^\prime 35^{\prime\prime}$   & $-14^\circ 22^\prime 05^{\prime\prime}$    & $-14^\circ 24^\prime 16^{\prime\prime}$       \\
Heliocentric systemic velocity (\kms) & 2330   &   2420       &  2500            \\
inclination (deg)                & 70          &   20         &   80          \\ 
Morphological type               & Sa pec      &  Sab         & Scd              \\
Corrected Blue total magnitude   & 12.23      &  12.51      &   16.95           \\
Absolute blue magnitude          & -20.23      & -19.95       & -15.51          \\
Blue luminosity (10$^{10}$ \Lo)  & 1.92        &  1.49        &  0.02           \\
FIR luminosity  (10$^{10}$ \Lo)  & 1.80        &  2.54        &     --           \\
\HI\ mass  (10$^{9}$ \Mo)          & 1.97        & 1.05         & 0.70             \\
\HH\ mass (10$^{9}$ \Mo)         & 1.19        &  0.39        &    --            \\
\Ha +\NIIt\ luminosity (10$^{40}$ erg/s) & 17        & 28           &     --             \\ 
20 cm emission (mJy)             & 180         & 45           &      --            \\
B-V color index (total,outskirts) & 0.79, 0.62 & 0.33, 0.40    &   0.46, --           \\
V-R color index (total,outskirts) & 0.51, 0.43 & 0.40, 0.39    &   0.30, --         \\
V-K$^\prime$ color index (total,outskirts)& 3.60, 3.03 &   --,--        &      --,--        \\
J-H  color index (total,outskirts)& 0.64, 0.41 &   --,--       &      --,--         \\
H-K$^\prime$ color index (total,outskirts)& 0.46, 0.47 &   --,--      &      --,--        \\
\tabsp \tableline \tabsp
\multicolumn{4}{l}{\parbox{\textwidth}{Sources: NED (positions, FIR fluxes), LEDA (morphology, corrected blue
magnitudes). The other data are from this paper. The velocity of FGC~0938 is the HI velocity. All optical magnitudes have
been corrected for  Galactic extinction. The magnitudes labeled ``outskirts'' have been measured with a polygonal
aperture that encompassed the outermost regions of NGC~2992/93, at a minimum radial distance of 30$\arcsec$ from the
nucleus and avoiding the prominent dust lane.}} \\
\end{tabular}
\end{table}

\clearpage

\begin{table}
\caption{Properties of the tidal features.\label{tab:tidal}}
\begin{tabular}{lccc}
\tableline\tableline \tabsp 
                       & NGC~2992 tail   &  NGC~2993 tail   &   Bridge     \\
\tabsp \tableline \tabsp 
Optical extent (kpc)   &  15.5           &  26.6            &    14.4      \\   
B magnitude  $^{a}$    &  14.63 \er 0.01 &  15.22 \er 0.01  &   15.51   \er 0.1     \\
B-V          $^{a}$    &  0.57  \er 0.02 &  0.39  \er 0.02  &   0.50:  \er 0.2       \\
V-R           $^{a}$    &  0.42  \er 0.02 &  0.45  \er 0.02  &   0.45:  \er 0.2      \\
\HI\ mass (10$^{8}$ \Mo) &  11.6           &  8.5             &   3.7        \\
Peak \HI\ column densities (10$^{21}$ \cmm) & 2.3           &  0.4              &   0.65        \\ 
V--band max Surface brightness (\sbu) & 22.1         &  23.3   &  24.0     \\
\tabsp \tableline \tabsp
\multicolumn{4}{l}{\parbox{\textwidth}{$^{a}$ Corrected for Galactic extinction}} \\
\end{tabular}
\end{table}

\clearpage

\begin{table}
\caption{Properties of the tidal object A245N.\label{tab:TDG}}
\begin{tabular}{ll}
\tableline\tableline \tabsp  
$\alpha$ (J2000)             &  $09^{\rm h} 45^{\rm m}  44.2^{\rm s}$                   \\
$\delta$ (J2000)              &  $-14^\circ 17^\prime 29.1^{\prime\prime}$                   \\
Heliocentric \HI\ velocity & 2175~\kms  \\
Mean optical velocity   &  2240~\kms                \\
Distance to progenitor  &  19.2~kpc                    \\
Size (D$_{24.5}$)         &  5.0 \x 10.5 kpc             \\
Blue magnitude   $^{a}$             & 15.21  \er 0.01                    \\
Absolute blue magnitude $^{a}$ &  -17.25   \er 0.01                 \\
Blue luminosity $^{a}$ & 12.3 \x 10$^8$ \Lo  \\
Blue central surface brightness $^{a}$ &  22.4 \sbu       \\
Optical/NIR colors  $^{a}$    &  B-V=0.55 \er 0.01 \\
                        &  V-R=0.41 \er 0.02,  V-K$^\prime$=3.14 \er 0.02  \\
                        &  J-H=0.32 \er 0.03, H-K$^\prime$=0.51 \er 0.03   \\
\HI\ mass                 &  8.9 \x 10$^8$ \Mo      \\
\HH\ mass                &    1.4 \x 10$^8$ \Mo     \\
\Ha +\NIIt\ luminosity          & \LHa = 0.74 \x 10$^{40}$ erg\,s$^{-1}$   \\
Star Formation Rate  $^{b}$   &  0.03 (0.13) \Mo\,yr$^{-1}$                 \\
Intrinsic extinction $^{c}$   & \Ab = 2.6 mag \\ 
\tabsp \tableline \tabsp
\multicolumn{2}{l}{\parbox{12cm}{$^{a}$ Corrected for Galactic extinction $^{b}$ Corrected for Galactic  and
  total extinction (values in parentheses) $^{c}$ Derived from the Balmer decrement towards TDG~4}} \\
\end{tabular}
\end{table}

\clearpage

\begin{table}
\caption{Spectrophotometry of  A245N.\label{tab:TDG:spec}}
\begin{tabular}{lcccc}
\tableline\tableline \tabsp 
                 & &     TDG~1$^{a}$           &   TDG~3$^{a}$  &   TDG~4$^{a}$   \\
\tabsp \tableline \tabsp 
\OIIt   & 3727      & --,-- & 183\er 23, 401\er 50 &  --,-- \\
\Hb     & 4861.3    & 100\er 37, 100\er 37  &  100\er 18, 100\er 18  & 100\er 17, 100\er 17 \\
\OIIIt  & 5006.9    & 35\er 28, 32\er 26  & 74\er 8, 68\er 7 &  88\er 17, 82\er 15 \\
\NIIt   & 6548.1    & 125\er 61, 51\er 25  &  --,-- & 80\er 18, 37\er 8 \\
\Ha     & 6562.8    & 705\er 172, 285\er 70  & 785\er 80, 285\er 29 & 622\er 64, 285\er 29 \\
\NIIt   & 6583.4    & 338\er 93, 135\er 37  & 290\er 33, 104\er 12 & 291\er 33, 132\er 15 \\
\SIIt   & 6716.4    & 200\er 68, 74\er 25  &  --,-- & 143\er 25, 60\er 11 \\
\SIIt   & 6730.8    & 101\er 41, 37\er 15 & --,-- &  79\er 16, 33\er 7 \\
\tabsp \tableline \tabsp
\multicolumn{2}{l}{Reddening Correction Factor (c)}        &  1.17: & 1.31  & 1.01 \\
\multicolumn{2}{l}{\Hb\ Equivalent Width~ (\AA)}        &   6  \er 2 & 9 \er 2 &  13 \er 2 \\
\tabsp \tableline \tabsp
\multicolumn{5}{l}{\parbox{\textwidth}{$^{a}$ First values correspond to the observed fluxes relative to
\Hb=100; second values correspond to the extinction--corrected relative intensities.}} 
\end{tabular}
\end{table} 

\clearpage

\begin{table}
\caption{\HI\ peak column densities in TDGs.\label{tab:TDGs:HI}}
\begin{tabular}{lccl}
\tableline\tableline \tabsp 
Object & \NHI\ & Resolution & Source \\
       & $10^{20} \cmm $  & kpc\,beam$^{-1}$  & \\
\tabsp \tableline \tabsp
NGC~2992/93 & 23        & 2.9 \x 2.1  & this paper \\
NGC~5291    & 2.2--12.6 &  7.3 \x 4.2 & Malphrus et al (1997) \\
NGC~7252    & 3.2--5.5  &  8.1 \x 4.8 & Hibbard et al (1994) \\
NGC~4038/39 & 2.3       & 4.9  \x 16.0 & van der Hulst (1979)  \\
Arp~105     & 3.2--9.6  & 13 \x 11.7 & Duc et al. (1997) \\
\tabsp \tableline
\end{tabular}
\nocite{Malphrus97}
\nocite{vanderHulst79}
\end{table} 

\clearpage

\end{document}